\documentclass[manuscript, screen, acmlarge, nonacm]{acmart}

\usepackage{natbib}

\usepackage[english]{babel}

\usepackage{amsmath}
\usepackage{graphicx}
\usepackage{lscape}
\usepackage{lipsum}
\usepackage{float}
\usepackage{hyperref}
\usepackage{ragged2e}

\begin{document}

\title{Understanding the Process of Human-AI Value Alignment}

\author{Jack McKinlay}
\authornote{Corresponding Author.}
\orcid{0000-0001-9822-8166}
\email{jam218@bath.ac.uk}
\affiliation{%
    \institution{University of Bath}
    \city{Bath}
    \country{UK}}

\author{Marina De Vos}
\orcid{0000-0003-3583-7671}
\email{cssmdv@bath.ac.uk}
\affiliation{%
    \institution{University of Bath}
    \city{Bath}
    \country{UK}}

\author{Janina A. Hoffmann}
\orcid{0000-0002-6246-2724}
\email{jah253@bath.ac.uk}
\affiliation{%
    \institution{University of Bath}
    \city{Bath}
    \country{UK}}

\author{Andreas Theodorou}
\orcid{0000-0001-9499-1535}
\email{andreas.theodorou@upc.edu}

\begin{abstract}
    {\bf Background:} 
    Value alignment in computer science research is often used to refer to the process of aligning artificial intelligence with humans, but the way the phrase is used often lacks precision.
    
    {\bf Objectives:}
    In this paper, we conduct a systematic literature review to advance the understanding of value alignment in artificial intelligence by characterising the topic in the context of its research literature. We use this to suggest a more precise definition of the term.
    
    {\bf Methods:}
     We analyse 172 value alignment research articles that have been published in recent years and synthesise their content using thematic analyses.
    
    {\bf Results:}
    Our analysis leads to six themes: value alignment drivers \& approaches; challenges in value alignment; values in value alignment; cognitive processes in humans and AI; human-agent teaming; and designing and developing value-aligned systems.
    
    {\bf Conclusions:}
    By analysing these themes in the context of the literature we define value alignment as an ongoing process between humans and autonomous agents that aims to express and implement abstract values in diverse contexts, while managing the cognitive limits of both humans and AI agents and also balancing the conflicting ethical and political demands generated by the values in different groups. Our analysis gives rise to a set of research challenges and opportunities in the field of value alignment for future work.
\end{abstract}

\maketitle

\section{Introduction}

\textit{Value alignment} is broadly understood as the challenge of ensuring that autonomous artificial agents act in ways aligned with humans and their values when deployed in society \cite[p.137]{russell2019human}. However, this is complicated by the complexity and diversity of human values, and their abstract, generalisable nature. 

There are hundreds of guidelines for how artificial intelligence (AI) systems should be developed and deployed, but values are too often described using non-specific language \citep{theodorou_towards_2020, alertubella_governance_2019}. This necessitates interpretation of these values by developers in the problem context while fulfilling the system's purpose. However, the subjective nature of this interpretation may result in inconsistencies in how values are achieved in different contexts, which is further complicated by how the opaque nature of many AI systems makes determining the influence of values in systems and their impacts difficult \citep{umbrello_mapping_2021, van_de_poel_embedding_2020}.

Values can be described as the drivers in individual and shared decision-making in humans: guiding factors shared across cultures \citep{schwartz1987toward}. A mutual understanding of each others' values enables us to cooperate on tasks despite differing backgrounds and interests. As autonomous agents become more prevalent as actors in society, their actions may not only realize human values, but also shape the development of these values \citep{cappuccio_can_2021}. To fully benefit from AI as a technology, it is essential to ensure that these autonomous agents act in accordance with our values so that their actions lead to desirable outcomes.

While researchers generally agree on the same high-level understanding of value alignment, definitions of value alignment presented in the literature are diverse and often shallow, if they are specified in the paper at all. Some authors take the view that value alignment should focus on coordinating objective functions \citet{sanneman_validating_2023}, which likely stems from the influential paper on cooperative inverse reinforcement learning by \citet{hadfield-menell_cooperative_2016}. Other authors like \citet{vamplew_human-aligned_2018, fisac_pragmatic-pedagogic_2020, lera-leri_towards_2022} provide a more high-level definition, which gives a broad goal but lacks direction in achieving it. \citet{serramia_qualitative_2020} provides a definition that seems far removed from the others in the literature, focusing on norm preference, yet is still explicitly referred to as value alignment. Given these diverse interpretations, papers with missing definitions \citep{bogosian_implementation_2017, peschl_moral_2022, siebert_estimating_2022} are problematic, as it makes it ambiguous how the authors have interpreted the problem of value alignment and how their work relates to it.

This inconsistent and imprecise terminology is understandable given the diversity of value alignment research. Yet, it makes it hard for researchers to develop a shared interpretation of the problem and poses challenges for researchers trying to assess the state of the field and identify useful research directions. The goal of this research is to provide a shared interpretation of the value alignment problem by analysing the different themes of value alignment research and to develop a conceptual model of value alignment as a process.

In order to develop our understanding of what value alignment is, we conduct a structured review of value alignment literature and analyse published articles using inductive thematic analysis. We take an inductive approach as this is a suitable approach when no prior theory to explain a concept has emerged \citep{naeem2023step}, while enabling us to build broader themes from participants' views \citep{braun2006using} and remain open to new themes that may emerge in the analysis \citep{fereday2006demonstrating}.

Our contributions in this paper are: 
\begin{itemize}
    \item a thematic analysis of 172 value alignment papers for the sake of extracting and analysing the core themes of the research;
    \item an in-depth analysis of the concepts and challenges in value alignment based on the identified themes;
    \item identification of valuable research directions within these themes based on the existing literature.
\end{itemize}

The rest of the paper proceeds as follows:
\begin{itemize}
    \item In Section \ref{sec:related_work} we discuss other works that analyse value alignment literature and compare their approach with our own.
    \item In Section \ref{sec:methodology} we explain our methodology.
    \item Section \ref{sec:analysis} contains our analysis of results.
    \item In Section \ref{sec:discussion} we discuss our the results of our analysis, directions for future research in the field, and limitations of our study. 
    \item Finally, Section \ref{sec:conclusion} concludes our analysis by highlighting the key observations from our analysis.
\end{itemize}

\section{Related Works}
\label{sec:related_work}

\citet{wallach_moral_2008} is one of the earliest reviews on embedding human values in artificial intelligence. The review provides a thorough introduction to the problem of learning values and distinguishes between top-down, bottom-up and hybrid ethical learning. 
Although technology has naturally developed since the book's publication, the ideas were cited frequently throughout our review, and hence can be expected to underpin the outcomes of our own analysis.

\citet{tolmeijer_implementations_2021} conducted a thorough review of the field of machine ethics, which value alignment has a close relationship with, and focused on machine ethics implementation. They produced a trimorphic taxonomy of the literature based on: i) the ethical theory included; ii) non-technical aspects; and iii) technical aspects. While the work contributes a means to categorise and analyse broad aspects of the field using this taxonomy, our approach instead focuses on identifying themes found across papers to produce a practical understanding of value alignment as a process without focusing on individual ethical theories. This leads to our synthesis of literature into a conceptual process, a distinct contribution from the classification taxonomy in the paper by \citet{tolmeijer_implementations_2021}.

\citet{heyder_ethical_2023} conducts a theoretical review on humans and AI agents interacting ethically in socio-technical systems for the sake of aligning values. They use this to produce a conceptual framework for the ethical management of Human-AI interactions. As in our review, they performed a qualitative analysis to identify patterns of themes that emerged in the reviewed articles. However, their review utilised a deductive approach with pre-existing codes, and explicitly analysed the work through a lens of \textit{sociomateriality}: the merging of technology and social phenomena in the emergence and observation of technology \citep{heyder_ethical_2023}. We instead use an inductive approach, and while our analysis incorporates many elements of agent-based approaches, our interpretation avoids using any specific theoretical framework.

\citet{zoshak_beyond_2021} produced another survey paper on artificial moral agents, concluding that there was a dominating presence of Western ethical values in the design of these agents, at the expense of other theories. Similar to \citet{heyder_ethical_2023}, they also used a deductive qualitative analysis approach, and similar to our paper, they used thematic synthesis to explore their results. Their work focuses on the normative aspects of value alignment, like \citet{tolmeijer_implementations_2021}. This leads to a strong focus on ethical theories that is absent from our own work. 

\section{Methodology}
\label{sec:methodology}

For our research we performed a structured literature survey and qualitatively coded the content of the included research articles. Qualitative coding reduces ``the data into meaningful segments and assigning names for the segments.'' \cite[p.183]{alma991003827159502761}. In this case, the data was the content of the included papers. A full description of our search terms, as well as the justification behind their use, is available in Appendix \ref{sec:search_details}.

The literature search was conducted in the Scopus database. We limited ourselves to English-language papers due to a lack of translation capability. We chose our search terms to find papers relating to value alignment and human preferences. After the relevance of topics including virtue ethics, the social contract \citep{rousseau2016social}, and multi-agent systems, became apparent in the initial returned papers, we performed an additional search with keywords on these topics to find relevant additional papers. We filtered on publication year, up to November 2023 inclusive, and subject area, only including papers tagged as computer science. We filtered on computer science to reduce the number of irrelevant papers from engineering or mathematics that, despite sharing similar themes with our exclusionary keywords, had not been removed from the search results due to the broad scope of the term "value alignment", which was still matching these papers. This did not exclude papers tagged with multiple disciplines including computer science. We also filtered for final publication papers only. In total, our search returned 734 papers.

We selected papers to be included in the survey based on our inclusion/ exclusion criteria. Our initial screening of papers to code had two phases: just the abstract and title in the first phase, and then a summary pass of the full text in the second phase. When inspecting the abstract and title, we excluded papers that used values in a non-socio-ethical sense, and we included those discussing incorporating human values or preferences into technology. When we investigated the full text of papers passing this stage, we included those that both attempted to define value alignment and discussed the challenges of getting humans and autonomous agents to behave in aligned ways, but excluded articles that only met one of these two criteria.  The inclusion/exclusion criteria are fully specified in Appendix \ref{sec:incl_excl_crit}. 

We performed an initial coding of the abstracts, introductions, and conclusions of the selected papers. This was done to generate initial codes quickly to frame our full coding in the later stage. During this process, we performed bibliography searches to add papers that were either relevant based on the cited information in the source paper or were highly cited across the papers under consideration. No further papers were added from the bibliographies of these additional papers in order to maintain a reasonable scope for this analysis.

From these initially coded papers, we applied a second set of inclusion/exclusion criteria that focused on including implementation-based papers over those focusing on the governance of AI or the moral status of autonomous agents. While these topics are relevant to value alignment, we left them for another analysis. We also excluded papers focusing only on a specific value, as value-specific issues were not the focus of this analysis. Finally, we included papers that discussed alignment in a normative sense in order to broaden our coverage

We used \textit{NVivo} for our coding. Coding was undertaken by a single coder. Coding consisted of assigning segments of the text or diagrams in the literature to codes that described the content of the segment in the context of characterising value alignment. We also coded for academic subjects that appeared in the text to see what subjects had been referenced in value alignment. Segments were generally passages of text rather than individual sentences, in order to retain the surrounding context. As per our inductive approach, we started with no codes and generated new codes as needed for categorising segments of text that felt relevant to the value alignment problem. All content in the papers we reviewed was considered for potential coding, except for titles, section headings and bibliographies.

Throughout the coding process, we grouped codes into categories, and categories into themes, based on commonalities between their content. This allowed our understanding of value alignment to be synthesised from the common themes that appeared across the literature, supported by the work of different authors in the field.

\subsection{Search Results}

\begin{figure}[ht!]
    \centering
    \includegraphics[width=0.9\textwidth]{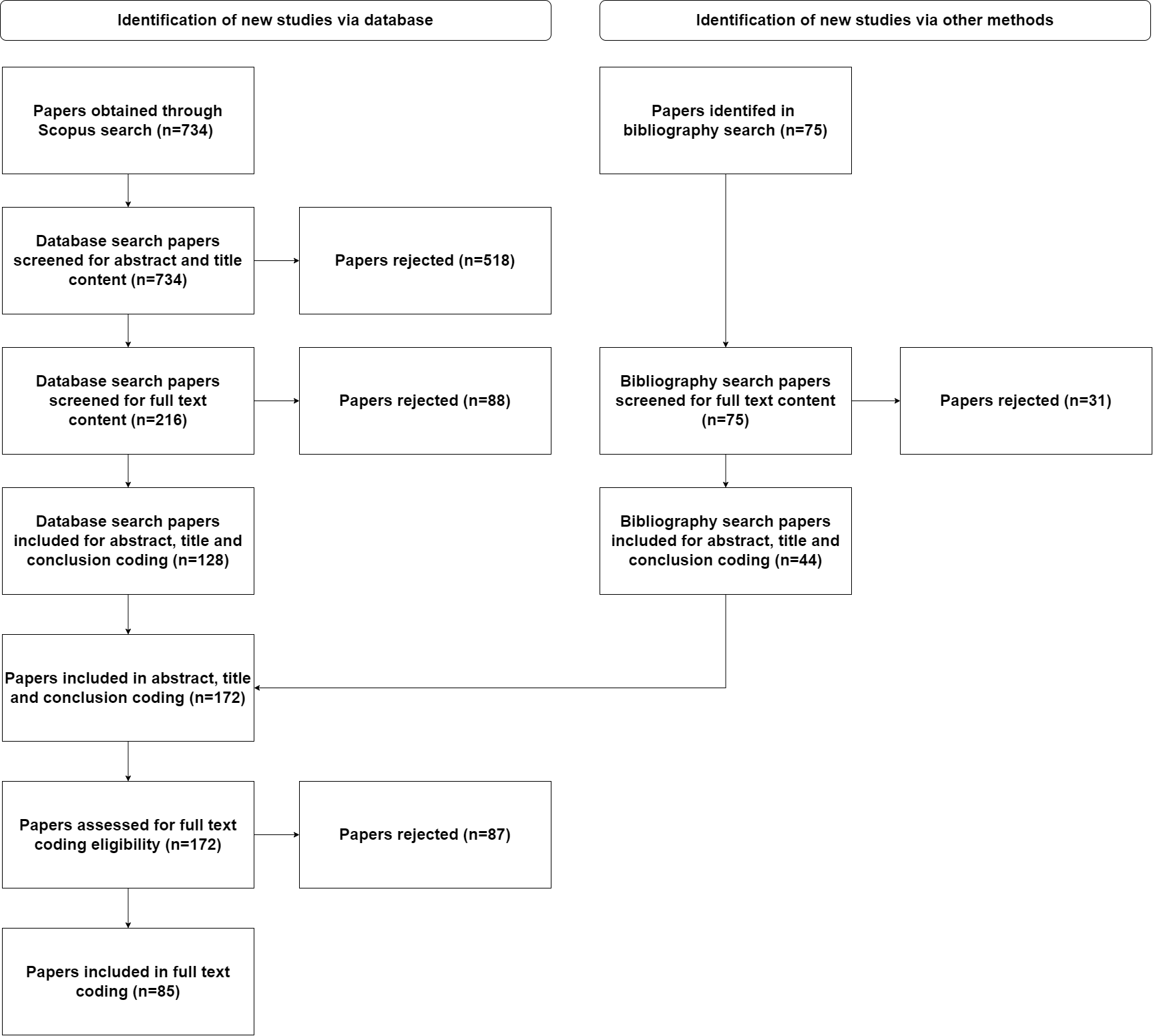}
    \caption{PRISMA flow diagram \citep{page2021prisma} illustrating our literature search results.}
    \label{fig:prisma_flow}
    \Description[From 734 papers in the initial search we arrived at 85 to code the full texts from.]{From our initial Scopus search result of 734 papers on the topic of value alignment, screening for abstracts and titles reduced this number to 216. Further screening the full text content reduced this to 128 papers, which had their titles, abstracts and conclusions coded. From here we added an additional 44 papers from the bibliographies of these 128 papers, giving us 172 in total. Screening the full text in these papers resulted in 87 rejections, resulting in 85 papers having their entire texts coded.}
\end{figure}

Fig~\ref{fig:prisma_flow} illustrates the literature search and screening process in a PRISMA flow diagram. Our initial search and screening resulted in 128 papers in which we coded the abstracts, titles and conclusions. 13 papers were omitted from consideration due to not being available in English. Searching the bibliographies from these papers led to the inspection of 75 additional papers, of which 44 were deemed relevant enough to include in the coding based on the full-text summary pass discussed in the previous paragraph. No further papers were added from the bibliographies of these additional papers in order to maintain a reasonable scope for this analysis. This resulted in 172 papers having their abstracts, introductions and conclusions used in coding. Our second set of screening criteria excluded 87 papers, leaving us with 85 papers to fully code. 

While reviewing the literature, we categorised papers based on the content, creating our own categories. This was done so that we could observe the types of literature being published and how this changed over time. 
\begin{itemize}
    \item Extended abstracts were short proposals of future work. 
    \item Research proposals were longer versions of the same and suggested several future agendas.
    \item Reviews examined previous literature.
    \item Theory proposals were arguments for ways of thinking about the value alignment problem.
    \item Methodology-focused papers were based on the development of a specific value alignment technique.
\end{itemize}

\begin{figure}[ht!]
    \centering
    \includegraphics[width=\textwidth]{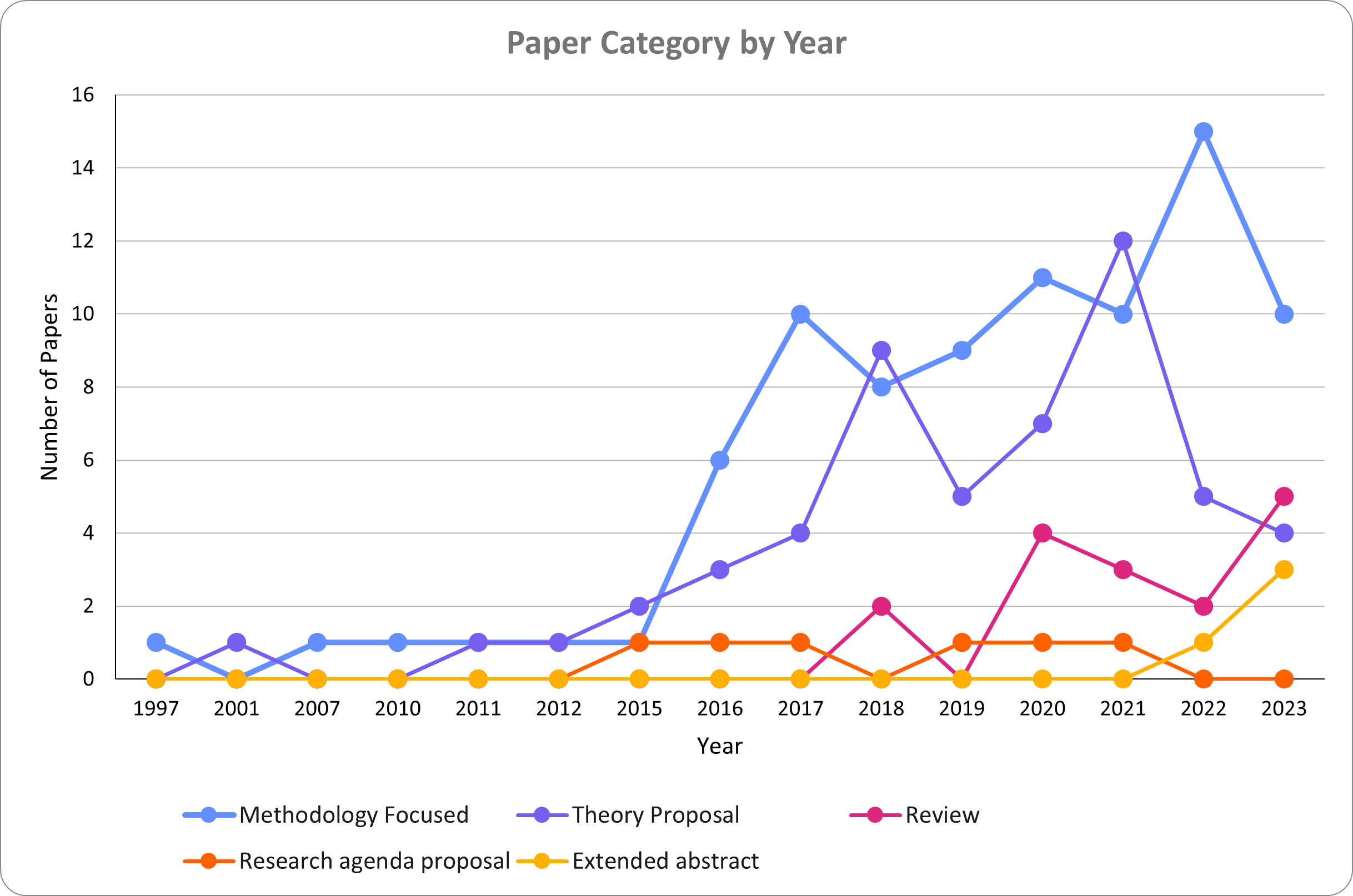}
    \caption{Timeline of papers included in our survey, grouped by the category assigned to the paper. Categories are explained in Section.~\ref{sec:methodology}.}
    \label{fig:paper_timeline}
    \Description[The types of papers submitted changed over time.]{From 1997 to 2023 we saw an increase in value alignment papers being published, with theory proposals and methodology focused papers increasing until 2022 before dropping in number. Review and extended abstract papers become more common after 20222. Research agenda proposals remained consistent at one per year between 2015 and 2021.}
\end{figure}

The categorised papers in the review are charted in Fig.~\ref{fig:paper_timeline}. We see that the field started publishing in earnest around 2015, with a generally increasing trend over time. To date the field largely consists of developing methodologies and proposals for how to think about value alignment issues, with research proposals for next directions appearing on a mostly-annual basis. The split between theory and methodology varies by year, but overall it would not seem to suggest any particular trends. In later years, we see reviews emerge as an increasingly popular category, suggesting increased attention on the field.

\section{Analysis}
\label{sec:analysis}

Our coding of value alignment literature identified six themes that characterise value alignment in AI. Within these themes, we identified categories that describe the majority of each theme's content. We list these, along with the three papers with the most excerpts coded for the given theme, in Table~\ref{tab:theme_breakdown}.

These themes group the codes generated from the analysed literature in the following way:

\begin{enumerate}
    \item \textbf{Value Alignment Drivers \& Approaches} focuses on understanding the motivations behind value alignment research, trends in approaches taken to studying the topic, and the range of subjects used in studying value alignment.
    \item \textbf{Challenges in Value Alignment} describes and discusses the identified challenges in accomplishing value alignment, in particular around how priorities are expressed and the challenges in reconciling different ethical systems.
    \item \textbf{Values in Value Alignment} highlights values themselves in value alignment research. It includes how values are embedded and represented in AI systems operating in diverse contexts, and how they change during a system's operation.
    \item \textbf{Cognitive Processes in Humans and AI} covers discussions on how humans use values in decision-making, including learning values and how they should be contextualised. It also discusses how these processes can be replicated in autonomous agents.
    \item \textbf{Human-Agent Teaming} discusses systems containing both humans and autonomous agents. It includes how these agents interact, and how information such as values and perceived system state is communicated in such systems. 
    \item \textbf{Designing and Developing Value-Aligned Systems} groups information pertaining to the process of designing and implementing value-aligned systems in practice. This includes understanding the stakeholders of such systems and how the alignment of systems with stakeholder needs can be tested.
\end{enumerate}

\begin{table*}[ht!]
\centering
\caption{\label{tab:theme_breakdown} Table of themes, representative categories and top three most relevant papers by number of coded data samples from that theme. The representative codes were taken from the highest-level groupings of codes within the given theme. The presence of papers in multiple themes demonstrates the interrelations between these themes.}
\begin{tabular}{p{0.3\linewidth}p{0.3\linewidth}p{0.3\linewidth}}
\hline
\textbf{Theme}                                        & \textbf{Representative Code Categories}                                                                                                                                                             & \textbf{Top Papers by Samples-in-Theme}                                                                                                                                                                \\ \hline
Value Alignment Drivers \& Approaches & \begin{tabular}[c]{@{}l@{}} Motivations\\ Technical and \\ Normative Alignment\\ Interdisciplinary Approach\end{tabular} & \begin{tabular}[c]{@{}l@{}}1. \shortcites{han_aligning_2022}\citet{han_aligning_2022} \\ 2. \shortcites{sutrop_challenges_2020}\citet{sutrop_challenges_2020} \\ 3. \shortcites{stenseke_artificial_2023}\citet{stenseke_artificial_2023}\end{tabular}            \\ 
Challenges in Value Alignment                         & \begin{tabular}[c]{@{}l@{}}Expressing Priorities\\ Implementing Ethical Theory\end{tabular}                                                                     & \begin{tabular}[c]{@{}l@{}}1. \shortcites{bench-capon_ethical_2020}\citet{bench-capon_ethical_2020} \\ 2. \shortcites{stenseke_artificial_2023}\citet{stenseke_artificial_2023} \\ 3. \shortcites{vamplew_human-aligned_2018}\citet{vamplew_human-aligned_2018}\end{tabular} \\ 
Values in Value Alignment               & \begin{tabular}[c]{@{}l@{}}Context \\ Value Dynamism \\ Value Aggregation \end{tabular}                                            & \begin{tabular}[c]{@{}l@{}}1. \shortcites{noriega_design_2022}\citet{noriega_design_2022} \\ 2. \shortcites{han_aligning_2022}\citet{han_aligning_2022} \\ 3. \shortcites{liscio_what_2022}\citet{liscio_what_2022}\end{tabular}                    \\ 
Cognitive Processes in Humans and AI                  & \begin{tabular}[c]{@{}l@{}}Learning\\ Reasoning \& Decision-Making\\ Self-Reflection in AI\end{tabular}                                                                                  & \begin{tabular}[c]{@{}l@{}}1. \shortcites{allen_artificial_2005}\citet{allen_artificial_2005} \\ 2. \shortcites{cervantes_autonomous_2016}\citet{cervantes_autonomous_2016} \\ 3. \shortcites{stenseke_artificial_2023}\citet{stenseke_artificial_2023}\end{tabular}     \\ 
Human-Agent Teaming                                   & \begin{tabular}[c]{@{}l@{}}Sharing Knowledge \\ Perception and Modelling\\ Interactions Between Humans \\ and AI\end{tabular}                                     & \begin{tabular}[c]{@{}l@{}}1. \shortcites{cappuccio_can_2021}\citet{cappuccio_can_2021} \\ 2. \shortcites{van_de_poel_embedding_2020}\citet{van_de_poel_embedding_2020} \\ 3. \shortcites{sanneman_validating_2023}\citet{sanneman_validating_2023}\end{tabular}     \\  
Designing and Developing Value-Aligned Systems        & \begin{tabular}[c]{@{}l@{}}Design \& Development Process\\ Stakeholders\\ Value Alignment Testing\end{tabular}                                                                          & \begin{tabular}[c]{@{}l@{}}1. \shortcites{noriega_design_2022}\citet{noriega_design_2022} \\ 2. \shortcites{gabriel_artificial_2020}\citet{gabriel_artificial_2020} \\ 3. \shortcites{brown_value_2021}\citet{brown_value_2021}\end{tabular}              \\ \hline
\end{tabular}
\end{table*}

\begin{figure}[ht!]
    \centering
    \includegraphics[width=\textwidth]{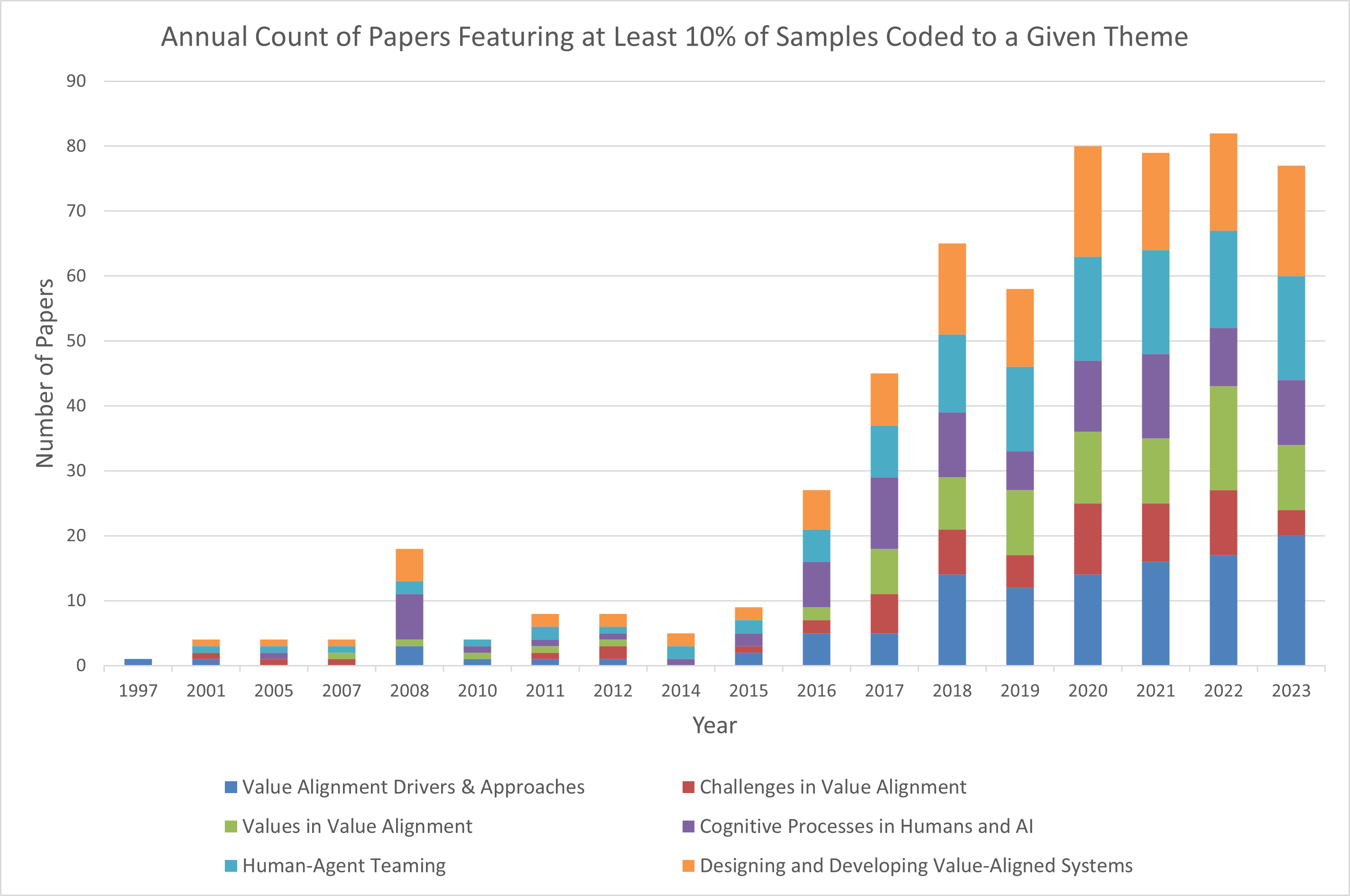}
    \caption{Number of studies in our survey, grouped by theme, that contained at least 10\% of its data samples coded to that group's theme, summed by year. We see that the representation of each theme remains broadly consistent, suggesting that these themes are characteristic of value alignment research.}
    \label{fig:StudiesByTheme_timeline}
    \Description[The proportion of papers in each theme remained consistent between 1997 and 2023.]{The ratios between the number of papers with at least 10\% of their codes attributed to a given theme remains broadly consistent from the start of the survey to the end. There is a shortage of papers before 2015, but after publication count increases the ratio between themes is consistent.}
\end{figure}

To demonstrate the distribution of themes in the literature, we created a chart (Fig.~\ref{fig:StudiesByTheme_timeline}) showing the number of papers from our survey that focused on each theme (with at least 10\% of samples coded to that theme) over time. We require at least 10\% of the samples to be coded to a given theme to include it in that theme's total.  This is to exclude papers which may have one or two data samples coded to a given theme, but this is negligible compared to the more prominent themes discussed in the paper. Over time, the percentage of papers mentioning each theme has remained broadly consistent. This indicates that these themes are not just passing trends but are core open challenges for value alignment research. While the 10\% threshold was chosen arbitrarily, testing other thresholds between 5 and 20\% did not significantly impact the resulting distribution of themes.

We now analyse the themes and the literature they describe in greater detail. We have structured our analysis to best illustrate value alignment given our investigation. This results in the first three themes having their own dedicated subsections. For the themes of cognitive processes and human-agent teaming, discussing them together as part of the value alignment process in systems proved more effective. Given that cognitive processes describe components internal to the agent, while human-agent teaming looked at interactions between the two agents, there were naturally significant interactions between the two themes, and we felt these were better portrayed in tandem than individually. Finally, our analysis led to many observations about the design and development of value-aligned systems across the themes, and we choose to highlight these where appropriate in other themes.

\subsection{Value Alignment Drivers \& Approaches}
\label{sec:Drivers_Approaches}

To characterise value alignment, it is essential to understand its motivation and any of its previous approaches. This reveals key objectives, concerns, and insights that have shaped the topic.

\subsubsection{Motivations}
\label{sec:motivations}

\begin{figure}[h]
    \centering
    \includegraphics[width=0.5\textwidth]{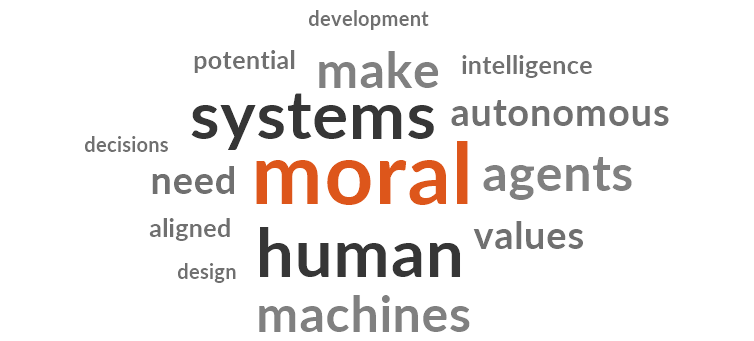}
    \caption{Word cloud of the 15 most frequent terms used in discussing the motivation to create value-aligned agents, grouping synonyms. We observed that \textit{autonomous} as a grouped term had the highest weighted percentage across terms, ignoring terms that described system aspirations (e.g. moral; values; aligned), or descriptors of system components (e.g. systems; human; machines).}
    \label{fig:motives_cloud}
    \Description[The most prominent words used when discussing value alignment were `moral', 'systems', and 'human']{'Moral' is the most prominent term in given value alignment motivations in the literature, followed by 'systems' and 'human'. Afterwards 'make', 'autonomous', 'agents', 'values', and 'machines' are the next tier of prominence. The less frequent terms out of the top 15 are 'development', 'potential', 'intelligence', 'decisions', 'aligned', and 'design'.}
\end{figure}

To better understand motivations for research on value alignment, we inductively coded our sample excerpts of given motivations, aiming to find common issues between texts. To support our analysis, we generated a word cloud of the 15 most prominent terms used in literature excerpts discussing motivations, grouping synonyms, as shown in Fig.~\ref{fig:motives_cloud}.

Combining this word cloud with an analysis of the text samples discussing motivation, we judged \textit{autonomy} to be one of the most prominent drivers for researchers' concern with the value alignment problem. \citet{vamplew_human-aligned_2018} summarises the concern driven by autonomy as:

\begin{quote}
    Increases in the agent's intellectual capacity, the broadness of the actions available to it, and the breadth of the domain in which it is applied increase the difficulty in ensuring the agent's behaviour is aligned, and also the magnitude of the negative side-effects of any unaligned behaviour.
\end{quote}

The autonomy driver can be broken down into several highlighted risks that were dominant in the literature. These included risks from \textit{unpredictability}, \textit{corrigibility}, and \textit{negative impacts on human values} through their interactions with autonomous agents.

Unpredictability as a risk stemmed from what \citet{yudkowsky_ai_2016} described as the problem of ``unforeseen instantiation'': humans' inability to properly anticipate all potential scenarios an autonomous agent may find itself in. As we endow autonomous agents with greater autonomy, the range of states they can explore, and the paths between these states, quickly become intractable to consider. Further complicating this is how an autonomous agent's view of the world is fundamentally different to our own. This exacerbates the autonomy risk by making it difficult to fully envision all the value-based risks from granting autonomous agents more ability to act independently.

Prominent within samples regarding the risk from unpredictability was concern regarding autonomous agents learning unintended strategies for obtaining rewards. This concept of \textit{reward hacking}, where agents learn to exploit loopholes in their reward functions to achieve greater rewards, is regularly cited as a concern in the alignment literature \citep{sanneman_transparent_2023, zhuang_consequences_2020}. Again, this difficulty in anticipating how an agent may learn to solve a problem obscures the potential value impacts. Taken together, unpredictability not only complicates autonomous agent design and training it also obscures the scale of risk that misalignment presents.

The idea of incorrigibility, introduced by \citet{soares2015corrigibility}, suggests that a sufficiently intelligent agent may not respond to attempts to correct its behaviour or shut it down, if it models doing so as counterproductive to maximising its utility function. If such an agent were to be misaligned with its stakeholder's values, then incorrigibility would pose a serious risk. This links to the concern around \textit{superintelligence}, as conceived by  \citet{bostrom2017superintelligence}. This was a frequently raised concern when discussing risk in value alignment, to the point where Bostrom was cited by approximately 25\% of the papers included in the first coding pass of our review. As \citet{arnold_value_2017} observed, the threat of uncontrollable AI agents strongly influenced the development of alignment initiatives. 

\citet{sarma_ai_2018} made the argument that there was no reason researchers could not pursue both near-term concerns and longer-term possibilities such as superintelligence, and that a continuum exists between these issues in the development of autonomous agents. The risk from misalignment comes from an agent's impacts on outcomes, rather than its capacity for cognition. So while the notion of superintelligence has had a prominent influence on motivating value alignment and corrigibility concerns, it should not be seen as fully, or even largely representative of the problem. 

Autonomous agents might also alter the values humans hold. \citet{han_aligning_2022} points out that it is nearly impossible to fully anticipate how technology will influence human actions, and the effect this will have on a system of values held by humans. \citet{svegliato_ethically_2020} also acknowledges this concern regarding the unpredictable effects AI systems can have on stakeholder values, which they state stems from performing gradual modifications to an agent's reward in the pursuit of achieving ethical behaviour. The idea of \textit{moral deskilling}, that extended interaction with autonomous agents may harm our morals as humans, is the focus of \citet{vallor_moral_2015}. Vallor concludes in the article that the moral future of humans interacting with emerging technologies is not sufficiently understood, and greater attention needs to be paid to this risk. With values changing over time and between contexts, which we illustrate in Section~\ref{sec:val_dyn}, and the need to aggregate values, discussed in Section~\ref{sec:val_agg}, it is evident that the values empowered by autonomous agents risk biasing human values in an undesirable direction or reinforcing outdated values, should misalignment occur.

We also noted in our analysis that the way AI has become, and continues to become, \textit{embodied in society} was almost as prominent a driver for value alignment work as autonomy. The presumed adoption of AI systems in decision-making capacities in society, and the delegation of higher-stake decisions to these autonomous agents, was rarely treated as a possibility rather than a certainty. \citet{sison_neo-aristotelian_2023} described this as an `inevitability' argument, though they argued against how truly `inevitable' the need to use autonomous agents in moral roles was, as well as the `inevitability' of humans choosing to develop autonomous moral agents. \citet{behdadi_normative_2020} noted that the growing range of industries that autonomous agents were being deployed, combined with their growing autonomy, increased the urgency of assessing the status of autonomous agents as moral agents. \citet{dignum_ethics_2018} made a similar observation but noted that the urgency was more in ensuring that autonomous agents acted in alignment with our ethics and values to build trust and ensure humanity's benefit, a sentiment echoed by \citet{bench-capon_ethical_2020} but with an emphasis on harm rather than trust.

A less mentioned risk in given motivations, but one worth highlighting, was the political risk from AI. As \citet{han_aligning_2022} observes, the development of AI by a small group of actors risks a kind of ``moral paralysis'' when the values embedded in a sufficiently powerful AI are determined by a small group of actors, and will have significant impacts on those who don't share these values. \citet{soares_value_learning_2018} points out that a sufficiently powerful system not only risks this moral paralysis, but also concentrates enormous power to influence the future of humanity into a small group of actors. The normative component of value alignment discussed in the next section, regarding which values should be built into the AI and who gets to decide, does little to help this concern.

Other common drivers for value alignment research included: a need for autonomous agents to act ethically \citep{svegliato_ethically_2020, badea_morality_2022}, and common values cited by policymakers such as safety \citep{juric_ai_2020, chaturvedi_ai_2023}, trust \citep{dignum_ethics_2018, firt_calibrating_2023}, and fairness \citep{osoba_steps_2020}. \citet{hagendorff_ethics_2020} also provides a further list of risks from autonomous agents that they considered neglected by modern guidelines for developing such systems.

Taking these motivations together, we can surmise that the objective of value alignment research is to manage the risks associated with autonomous agents (unpredictability, corrigibility, and impacts on human values from human-AI interactions) in such a way that they can be allowed to impact our values in a positive manner, while also accommodating the inherent political and ethical risks associated with value influence.
 
\subsubsection{Technical and Normative Alignment}
\label{sec:tech_and_norm}

Value alignment is often described in terms of both normative alignment problems and technical alignment problems \citep{gabriel_artificial_2020, chaturvedi_ai_2023, sutrop_challenges_2020}. The former refers to the challenge of deciding what values and principles a system should be aligned with, while the latter refers to the challenge of getting the system to align with these values.  
\citet{firt_calibrating_2023} also suggested a third dimension, calibration problems, which focuses on fine-tuning systems for specific values for trust building. However, they presupposed in their discussion that this would become relevant after the first two problems were solved.

While this separation between normative and technical is convenient for narrowing the scope of a given research paper, it does obscure the interaction between these two sides of the same coin. As \citet{tolmeijer_implementations_2021} states, ``In the context of implementing machine ethics, it can be a pitfall for philosophers to use a purely theoretical approach without consulting computer scientists, as this can result in theories that are too abstract to be implemented. Conversely, computer scientists may implement a faulty interpretation of an ethical theory if they do not consult a philosopher.''. Additionally, \citet{peschl_moral_2022} observed that the value alignment field has largely focused on reward learning, a technical problem, as demonstrated by the prominence of \citet{hadfield-menell_cooperative_2016} throughout our analysed papers. Research exploring the normative side was comparatively rarer in our survey, mainly discussed at higher conceptual levels \citep{gabriel_artificial_2020, butlin_ai_2021} rather than developed into actionable processes. 

However, \citet{robinson_action_2023} describes this normative aspect as crucial to good alignment: ``The first step is to find principles that the AI agent possibly could and should be aligned with. There is no point trying to align AI agents to principles they cannot be aligned with, or that we have no interest in aligning them with.''. As \citet{sutrop_challenges_2020} cautions, AI developers risk assuming that a normative solution will naturally follow from sufficiently technically capable AI, even though we as a society still remain undecided about our value priorities. This also links into the need to aggregate our values across more groups and cultures than just the developers of such systems to enable alignment, which we discuss further in Section~\ref{sec:val_agg}. Focusing on the technical side of alignment, as was indicated in our analysis, at the expense of the normative elements risks neglecting important aspects of the process.

\subsubsection{Interdisciplinary Approaches}

\begin{figure}[h]
    \centering
    \includegraphics[width=1.0\textwidth]{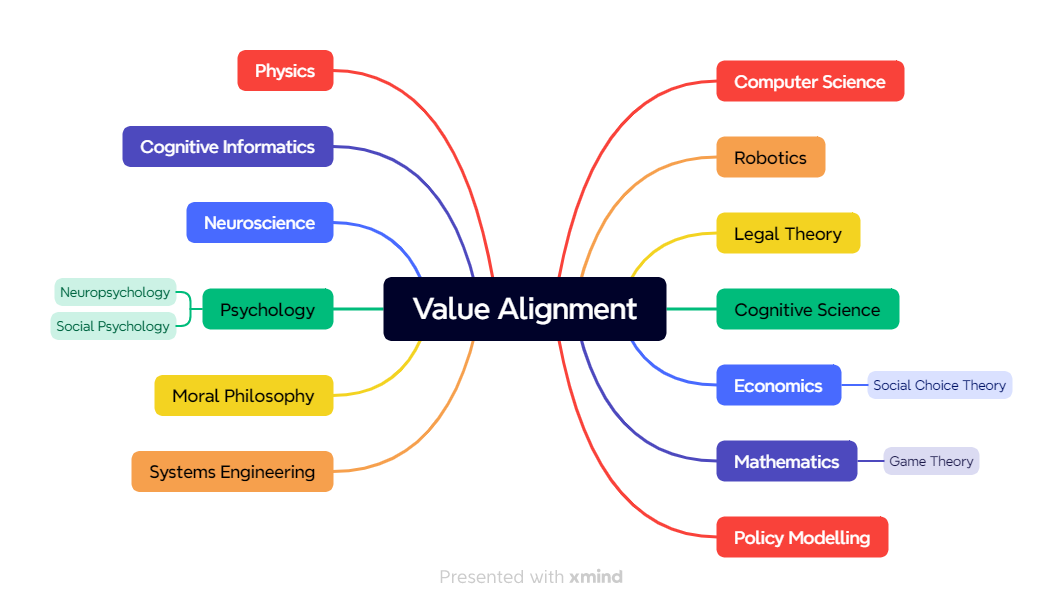}
    \caption{A non-exhaustive range of subjects previously applied to the value alignment problem. Note that many of the subjects applied, such as robotics, systems engineering and cognitive science, are themselves considered interdisciplinary. This expands the range of relevant subjects further.}
    \label{fig:ValAl_subjects}
    \Description[Disciplines appearing in value alignment research covered a range of subjects.]{The disciplines used in the value alignment literature included: computer science; robotics; legal theory; cognitive science; economics and the subtopic of social choice theory; mathematics and the subtopic of game theory; policy modelling; physics; cognitive informatics; neuroscience; psychology and the subtopics of neuropsychology and social psychology; moral philosophy; and systems engineering.}
\end{figure}

An important observation from examining previous value alignment work is that value alignment is often seen as an \textit{interdisciplinary} challenge, a sentiment echoed by authors including \citet{russell_research_2015, noriega_design_2022, sutrop_challenges_2020, dafoe_cooperative_2021, khan_ethics_2022}, and \citet{li_problems_2021}. Numerous subjects contributed to the different approaches across the value alignment research literature, which we illustrate in Fig.~\ref{fig:ValAl_subjects}. Many of these subjects were themselves interdisciplinary, further expanding the list of subjects relevant to the value alignment problem. This highlights both the importance and challenge of integrating a diverse body of knowledge into the research and development process of value-aligned systems.

The contributions of the subjects most frequently mentioned as being relevant to value alignment is well summarised by \citet{dafoe_cooperative_2021}: ``psychology, to understand human cognition; law and policy, to understand institutions; history, sociology and anthropology, to understand culture; and political science and economics, to understand problems of information, commitment and social choice.''. In further detail, we see psychology, neuropsychology and cognitive sciences suggested for understanding values and value structures in concept and in action \citep{butlin_ai_2021, sarma_ai_2018, cervantes_autonomous_2016, haas_moral_2020, montes_value_2023, ratoff_can_2021, fisac_pragmatic-pedagogic_2020}; philosophy and legal studies for understanding values, moral hazards and explainability \citep{zoshak_beyond_2021, hadfield-menell_incomplete_2019}; and humanities subjects applied to understanding human decision-making, cooperation and value aggregation \citep{holgado-sanchez_admissible_2023, lera-leri_towards_2022}.

The main barrier to interdisciplinary cooperation is translating ideas between domains. For example, \citet{cervantes_autonomous_2016} noted that emulating the brain's complex decision-making process in autonomous agents was enormously difficult, in part due to the lack of an accurate formalism of human brain functions that could be replicated in autonomous agents. This limits our ability to exploit neuropsychology results in AI. \citet{rahwan_society---loop_2018} discusses the cultural divide that exists between engineering and the humanities: while practitioners of the humanities are skilled at considering the applications of moral hazards, they can struggle to articulate them in ways that engineers can operationalise. Similarly, engineers are not always able to quantify their own work in a way that can be understood by humanities practitioners.  Given the need to integrate normative and technical aspects simultaneously, this cultural divide presents a significant barrier to doing so.

\subsection{Challenges in Value Alignment}
\label{sec:Chall_in_ValAl}

We have examined the difficult nature of doing value alignment in Section~\ref{sec:motivations}, given the opaque nature of autonomy and in quantifying the scale of value impacts. In this section, we expand on the particular challenges highlighted by our analysis from the perspective of an automated process involving interacting humans and autonomous agents.

\subsubsection{Expressing Priorities}
\label{sec:expressing_prios}

Throughout the literature, there were repeated efforts to make values known by both autonomous agents and humans. In the former case, this was predominantly in the form of these agents learning from humans, while the latter involved enabling humans to both inspect values and understand their own desires.

During our analysis, we observed that values, goals and preferences were used interchangeably as targets for alignment throughout the literature. This observation was also made by \citet{gabriel_artificial_2020} and \citet{zurek_value-based_2021}, with the former pointing out that these concepts are not equivalent. Our analysis would suggest that the terms can be interpreted as follows: 

\begin{itemize}
    \item Values, existing at the most general and abstract level, are used in forming desirable goals in more specific contexts and evaluating possibilities.
    \item Goals are formed from the values manifesting in the given context, with goals constructed in pursuit of those values. The values that have led to a goal being chosen will not necessarily be obvious when auditing the goals, however.
    \item Preferences can be understood in the conventional sense as priorities between different options and outcomes based on our underlying goals and values, including contextual priorities between goals and values themselves. 
\end{itemize}

Because values serve as expressions of our goals \citep{schwartz1992universals}, the challenges in expressing our goals are inherently tied to the value alignment problem. Because values serve as expressions of our goals \citep{schwartz1992universals}, challenges in expressing our goals consequently lead to challenges in expressing our values. Hence, the difficulty in goal expression is an essential issue in the value identification problem. \citet{hadfield-menell_incomplete_2019} describe the problems in expressing goals as being entirely responsible for the alignment problem. While we would not want to point to goal misspecification as the sole root of challenges in value alignment, as we find this too reductionist, it is certainly worth continued attention. 

Given that many of the goals we would ideally use AI agents for are complex or hard to define \citep{christiano_deep_2017}, it seems inevitable that we struggle to define these goals correctly or fully. For example, \citet{thornton_incorporating_2017} points out that, ideally, we want self-driving vehicles to drive safely and smoothly, but this requirement does not translate into machine language easily. \citet{hadfield-menell_incomplete_2019} also agrees that the misspecification of AI reward functions is often unavoidable. Many of the challenges in expressing goals can be described in terms of under-specifying goals \citep{alamdari_be_2022}, with our stated goals failing to capture important nuances. 

This underspecification often occurs in the form of developers attempting to consolidate goals into utility functions, a formula for capturing the preference of one outcome over another. However, these often fail to capture what we really desire \citep{aliman_augmented_2019}. As \citet{vamplew_human-aligned_2018} notes, maximising expected utility functions, the predominant paradigm in modern machine learning architectures, often leads to unexpected behaviour that fails to align with the designer's original goal. While a popular approach, innovation is still needed to consider alternative mechanisms beyond utility functions to encode our values.

Expressing values and preferences directly, circumventing goals in the process, is also a challenge. Humans lack adeptness at expressing their preferences in explicit statements \citep{bharadhwaj_auditing_2021, milli_should_2017, liscio_value_2023} or quantified forms \citep{rosenthal_monte_2012}, and struggle to mentally compare sufficient scenarios to cover all eventualities \citep{visser_interest-based_2011, rosenthal_monte_2012}. Preferences also vary between individuals, complicating preference modelling by autonomous agents due to the need to be able to adapt to differences in preferences, or to aggregate them \citep{rosenthal_monte_2012}. People face similar challenges in expressing \citep{sanneman_validating_2023} or evaluating \citep{sanneman_transparent_2023, liscio_what_2022} values, struggling to express values outside of concrete examples \citep{liscio_value_2023} as a result of their abstract, contextual, and often incommensurable nature.

Taken together, the difficulties in expressing values, goals and preferences illustrate one of the fundamental challenges in value alignment: expressing our values in the first place, either as values, goals, or preferences, is hard. Research explicitly attempting to improve how values, goals and preferences can be expressed was not found in the survey, presenting a possible opportunity for the future. If we could define our values, goals or preferences more accurately, this would help to reduce the noise in communicating these concepts to autonomous agents while generating evidence of what has been shared in the learning process, making alignment a smoother and more scrutable process. 

\subsubsection{Implementing Ethical Theories}
\label{sec:impl_ethic_theo}

The other most prominent challenge in the value alignment literature reviewed was the difficulty of summarising and translating the millennia of ethical thought into a machine-compatible format. Ethical systems serve as vital repositories of our developed values as groups of humans, and no good argument for not making use of them in aligning humans and autonomous agents emerged from the analysis. However, different systems of ethics present different frameworks for modelling values in autonomous agents and have implications for the alignment process.

Within the reviewed papers, the ethical discussion centred around three Western theories: consequentialism, usually in the form of utilitarianism; deontology; and virtue ethics. \citet{cappuccio_can_2021} provides a succinct summary of the three theories:

\begin{quote}
Consequentialism proposes to evaluate an action's moral 
value based on the utility created by the action's effects. Deontology, in turn, focuses on the intentions that motivated the action and judges whether or not the agent acted out of authentic goodwill to fulfil their obligations and respect others' rights. \ldots Virtue Ethics differs from both Consequentialism and Deontology because, unlike them, it does not primarily ask whether an action produces desirable effects or is motivated by good intentions, but whether a person deserves praise or blame and whether the kind of life they live is worth living. 
\end{quote}

Consequentialism and the related utilitarianism have thrived under machine learning paradigms. Credit is given to the popularity of utility functions and reinforcement learning \citep{gabriel_artificial_2020} and its pre-existing history with economics \citep{vamplew_human-aligned_2018}. \citet{franzke_exploratory_2022} also claims that utilitarian values pervade AI governance. As a result, deontology-based approaches have been left feeling marginalised, with most of the papers identified as deontic in our survey being theoretical analyses rather than implementations \citep{pagallo_even_2016, rahwan_society---loop_2018} --- though some deontic ideals seem to manifest in reasoning-based implementations \citep{cranefield_no_2017, szabo_understanding_2020}. Meanwhile, Virtue ethics has been championed as an under-represented \citep{murray_stoic_2017, vamplew_human-aligned_2018} but robust foundation for value alignment \citep{pagallo_even_2016, cappuccio_can_2021, franzke_exploratory_2022}, but concrete implementations of the framework remain in their infancy \citep{govindarajulu_toward_2019, crook_anatomy_2021, stenseke_artificial_2023}.

Even if priorities and ethical theories are able to be successfully expressed and implemented, there remains the design challenge of choosing between them. As we mentioned in Section \ref{sec:tech_and_norm}, technical value alignment work often ignores the normative aspect. However, in choosing (implicitly or explicitly) an ethical theory and the goals and values for a system to align to, a value-laden decision is itself being made about which values and goals are worth empowering through artificial intelligence systems \citep{davoust_social_2020}. As stated by \citet{soares_value_learning_2018}, ``human goals are complex, culturally laden, and context-dependant'', demonstrating how our choice of goals incorporates our own cultural biases. Given the political risks of artificial intelligence discussed in Section \ref{sec:motivations}, these decisions must be made carefully if we want to avoid marginalising schools of ethical thoughts or particular values, particularly given the dominance some cultures currently enjoy in AI development.

Comparisons between ethical theories pervade the literature. To quote \citet{bench-capon_ethical_2020}: ``it is seen that each approach has its own particular strengths and weaknesses when considered as the basis for implementing ethical agents, and that the different approaches are appropriate to different kinds of system.''. We list the strengths and weaknesses that we identified for different frameworks in the surveyed literature in Table \ref{tab:ethics_compare}. 

\begin{table*}[ht!]
    \centering
    \caption{List of the strengths and weaknesses of consequentialism and utilitarianism, deontology, and virtue ethics discussed in the value alignment literature analysed in our survey.}
    \begin{tabular}{p{3cm} p{5cm} p{5cm}}
        \hline
        \textbf{Ethical Theory} & \textbf{Strengths} & \textbf{Weaknesses} \\ 
        \hline Consequentialism \& Utilitarianism & Mathematically robust \citep{stenseke_artificial_2023}; efficient to implement \citep{bench-capon_ethical_2020}; transparent reasoning \citep{bench-capon_ethical_2020}; encodes existing knowledge \citep{bauer_virtuous_2020}; gives precise guidance \citep{murray_stoic_2017}  & Struggles with incommensurable ethical objectives \citep{allen_artificial_2005, eckersley_impossibility_2019}; prone to reward hacking \citep{sanneman_transparent_2023, christiano_deep_2017}; computational complexity \citep{allen_artificial_2005, bench-capon_norms_2017}; difficult to comprehensively assign utility to actions/outcomes \citep{mechergui_goal_2023, thornton_incorporating_2017, yudkowsky_ai_2016}; doesn't explain human behaviour well \citep{atkinson_value_2016, gavidia-calderon_what_2022}\\
        Deontology & Straightforward to implement \citep{stenseke_artificial_2023, bench-capon_ethical_2020}; supports cooperation \citep{pagallo_even_2016}; gives precise guidance \citep{murray_stoic_2017} & Requires sufficient background knowledge for interpreting rules \citep{allen_artificial_2005, badea_morality_2022}; creates normative conflicts \citep{bench-capon_ethical_2020, bauer_virtuous_2020}; dependant on good rules \citep{bench-capon_ethical_2020, pagallo_even_2016, murray_stoic_2017}; rigid \citep{constantinescu_understanding_2021, bauer_virtuous_2020}\\
        Virtue Ethics & Supports flexible contextualisation of virtues \citep{franzke_exploratory_2022, pagallo_even_2016}; supports development of ethical behaviour in self and others \citep{coleman_android_2001, reichberg_applying_2021, stenseke_artificial_2023, cappuccio_sympathy_2020}; supports embodiment of artificial agents in society \citep{gamez_artificial_2020}; supports explainability \citep{bench-capon_ethical_2020}; supports hybrid ethical learning \citep{liu_human---loop_2022} & Very difficult to implement \citep{stenseke_artificial_2023, sutrop_challenges_2020, govindarajulu_toward_2019}; difficult to define virtues \citep{cappuccio_can_2021}; lack of uniform virtues \citep{pagallo_even_2016}; lack of specific guidance on ethical behaviour \citep{murray_stoic_2017} \\ \hline
    \end{tabular}
    \label{tab:ethics_compare}
\end{table*}

Given the individual strengths and weaknesses of different approaches, an instinctual response might be to assume that no single theory can suit all value alignment needs. Such an observation has been made by several authors \citep{pagallo_even_2016, vamplew_human-aligned_2018, cappuccio_can_2021}. As \citet{gabriel_artificial_2020} puts it: ``it is very unlikely that any single moral theory we can now point to captures the entire truth about morality. Indeed, each of the major candidates, at least within Western philosophical traditions, has strongly counterintuitive moral implications in some known situations, or else is significantly underdetermined.''. Given the role of ethical systems in representing groups of values, we can infer from this statement that no single ethical system used to perform value alignment would suit all possible situations.

Considering the inadequacy of a single moral theory, the notion of combining ethical frameworks has gained popularity within the value alignment literature. Drawing from \citet{allen_artificial_2005}'s initial rationale for hybrid ethical learning systems, researchers acknowledge the merits of combining ethical frameworks to alleviate the weaknesses of individual ethical theories. Some practical examples of hybrid ethical learning systems can be seen in \citet{thornton_incorporating_2017, cordova_practical_2022} and \citet{cervantes_autonomous_2016}. However, \citet{stenseke_artificial_2023} cautions about cherry-picking in this approach, where suitable features are chosen from ethical theories without considering them in the context of holistic ethical cognition, as it risks neglecting to consider how these features relate to the wider reasoning of agents in complex dynamic contexts.

On a final note, \citet{svegliato_ethically_2020} suggests that a chosen ethical framework may be incompatible with the goals of the system. Rather than being seen as a failure of the system, it should instead draw the designer's attention to the ethics of the task itself, and whether it is worth pursuing. \citet{davoust_social_2020} calls this a social dilemma: a conflict between what is good for the wider group and what the autonomous agent (or its owner) wishes to achieve. In this sense, ethical frameworks, while not always simple to work with, can serve as important ethical sense checks for developing value-aligned systems. To develop ethical, value-aligned agents, the choice of an ethical framework should be guided by aspirations rather than practicalities. To ensure accountability and transparency in this process, the design decisions should be thoroughly documented, including a clear justification for the selected ethical system(s).

Summarising, ethical schools of thought are a highly relevant topic in value alignment, but their different aspects present challenges for both selection and implementation. At the same time, relying too much on a single framework risks leaving its particular weaknesses unresolved and the advantages of another framework unused. This challenge of exploiting and combining ethical frameworks, already used to structure our daily society's functioning, is a key obstacle in making use of our existing value-based knowledge in the creation of autonomous agents that can integrate with said society.

\subsection{Values in Value Alignment}
\label{sec:vals_in_ValAl}

We now reach the central concept of value alignment as a topic. The abstract nature of values necessitates processing them before they can be used by digital agents. For example, translating goals and preferences discussed in Section~\ref{sec:expressing_prios} is one form of this processing. Here, we examine the nature of this value processing. 

It is important to establish that most reviewed papers based their interpretations of values on the work of \citet{schwartz1992universals, schwartz2012overview}. While some reference was made to other values scholars like \citet{rokeach1967rokeach} and his work on value classification, and the work by \citet{graham2013moral} on moral foundations theory, these were in the minority. It is easy to understand why this is the case, as Schwartz's theory is well supported by empirical evidence, and its structure lends itself easily to machine implementation. Regardless, the applicability of other interpretations of value systems in value alignment remains under-explored. 

As mentioned in the introduction, values can be understood as latent drivers in evaluating situations and outcomes. Possessing a combination of contextualised values can lead to positive, negative or indifferent assessments of a presented situation. This is a simplified characterisation, but it will support the remainder of this section.

\subsubsection{Stakeholders}
\label{sec:val_al_stake}

So far, we have discussed value alignment in the context of autonomous agents aligning with human values, making human stakeholders the primary sources of values that a system needs to align with. These values can occur from individuals, but in the literature, they were more often discussed as resulting from the aggregation of particular groups, such as developers or users of AI systems. While stakeholder groups have been identified (\citet{tomsett2018interpretable},\citet{meske2022explainable}), we found developers and users to be the main focus. 

Our analysis reveals that effective alignment requires a two-pronged approach: first, understanding a stakeholder's values so that they can inform an agent's actions, and second, empowering those values to ensure the agent adds value and avoids undermining existing ones. This relates back to both the relevance of an agent's influence on values from Section~\ref{sec:motivations}, and the difficulties in expressing values from Section~\ref{sec:expressing_prios}.

Understanding stakeholders was often framed in terms of understanding stakeholder values and the variance between stakeholders. The technical approach to this was primarily reinforcement learning and utility functions \citep{hadfield-menell_cooperative_2016, ficici_simultaneously_2008, zintgraf_ordered_2018}, but other methods such as CP-nets \citep{loreggia_metric_2019}, or multi-label classification with orderings over values \citep{siebert_estimating_2022} were also observed. Engineering approaches were also used, including requirement engineering from software engineering \citep{liscio_what_2022} and guided annotation \citep{liscio_collaborative_2021}.

\citet{liscio_value_2023} states that this stakeholder value inference process cannot be performed entirely through computational means, however, since behaviour alone may not reveal enough about values. \citet{gabriel_artificial_2020} also raises the question of autonomous agents following instructions versus implicit stakeholder needs, which brings back the issues in expressing goals for value alignment from Section~\ref{sec:expressing_prios}. \citet{siebert_estimating_2022} showed improved performance at estimating value rankings of users when combining stated motivations with choices made as information for the autonomous agent. Still, the gain was relatively small compared to the effect of only inferring over given motivations. This might suggest a difference between the values driving user choices and the values users report when asked to justify their choices, further complicating this inference process.

On the topic of empowerment, \citet{feffer_preference_2023} raises the imperative that the AI ethics community has to empower stakeholders of autonomous agent systems in the co-creation of such systems. The authors, along with \citet{siebert_estimating_2022}, call for a more participatory design of these systems. \citet{vallor_moral_2015} argues for a further form of this, desiring a cultural shift to bring all members of society as collectively responsible in the design process rather than leaving it in the space of product designers and marketers. \citet{alertubella_governance_2019} supports aspects of this notion, suggesting the value of educating the population about their ability to shape the development of society through responsible AI technology. \citet{liscio_value_2023} also points out the importance of ensuring stakeholder consent and transparency about the fairness of the process. While they discuss this in the context of value inference, it relates to the idea of stakeholder empowerment through making them active decision-makers in the process of interacting with these systems, rather than just sources of data. 
 
These papers indicate that empowerment is not just about designing autonomous agents to satisfy values, but also incorporating more stakeholders into the design process so that these values can be properly recognised in the first place. It should be understood that stakeholders in value alignment are not just sources of information about values, but active participants in the system aiming to achieve a value-aligned state.

\subsubsection{Contextualisation}
\label{sec:contextualisation}

The key process in transforming values from ``abstract motivations that guide our opinions and actions'' \citep{schwartz2012overview} to decision-making factors is \textit{contextualisation}. Contextualisation is the interpretation of values in a given context and representing them through contextual proxies so that they can be actioned by autonomous agents.

Values are contextualised by several means in the literature, including: 

\begin{itemize}
    \item Goals \citep{badea_morality_2022, cranefield_no_2017, montes_value_2023} consistent with the definition from \citet{schwartz1987toward} that values are grounded as goals.
    \item Norms \citep{alertubella_governance_2019, zurek_reasoning_2022, pigmans_role_2017} as representations of grouped values.
    \item Design features in the agent's environment that could be interacted with directly, or measures of some combination of these interactive features \citep{cranefield_no_2017, noriega_use_2022}, in line with the work by \citet{van2013translating} on embedding values through value hierarchies.
\end{itemize}

Value hierarchies in particular illustrate how contextualisation relates to the \textit{embedding} of values in autonomous agents through design. \citet{noriega_addressing_2023} outlines a generalised heuristic approach to contextualisation:

\begin{quote}
    [Online Institution (OI)] values can be contextualised and embedded in four successive stages: (i) values for the application domain and [Conscientious Design] categories for the consensual preferences of the three design stakeholders towards the OI, (ii) for the individual preferences of each of the design stakeholders of the OI; (ii) then for the compatibility requirements of the situated OI; and, finally, (iii) for the six WIT\footnote{W - World; I - Institution; T - Technology}-articulation design concerns (abstraction, grounding, specification, implementation, input and output).
\end{quote}

In this heuristic we see contextualisation, and in tandem context, described in terms of values, applications, stakeholders and institutions, and design constraints. Intuitively, a general context structure is very difficult to define, as understanding the influence of various factors on decision-making is a problem that has challenged decision scientists for decades \citep{spektor2021elusiveness}. That said, we concur with \citet{noriega_addressing_2023} that the concepts above are effective starting points to consider in approaching value contextualisation in practice.

\citep{van_de_poel_embedding_2020} also discusses this process of contextualisation, though not in explicit words. In this paper, the author emphasises the nature of embedded values in an autonomous agent and how the agent should adapt to different contexts to better realise these values. Here, we see values treated as something internal to the agent: intended objectives embedded through design that need to be approached differently depending on the context. This contrasts the idea of treating values as concepts held externally to an agent by stakeholders, which are interpreted by the agent to produce different objectives in relation to contextual proxies in the environment. This would effectively make the agent's only value satisfying its stakeholders, a flexible but unpredictable objective. Considering the risks associated with autonomy discussed in Section \ref{sec:motivations}, a safer design approach would be to embed clear, well-defined values and contextualization methods for the agent and its associated task. This would promote predictability and auditability.

We see from these papers that contextualisation is a process impacting both the design and ongoing operation of an autonomous agent. The sheer diversity of contexts and the fact that they are difficult to fully anticipate in advance make programming these in advance impractical, if not impossible, as we alluded to in Section~\ref{sec:motivations} when discussing autonomy. A means of expressing contexts accurately to autonomous agents in an ongoing fashion, an extension to the challenge of expressing values, and methods for autonomous agents to learn contexts on their own, would be essential for reliable value-aligned systems.

\subsubsection{Value Dynamism}
\label{sec:val_dyn}

\begin{figure}
    \centering
    \includegraphics[width=1.0\textwidth]{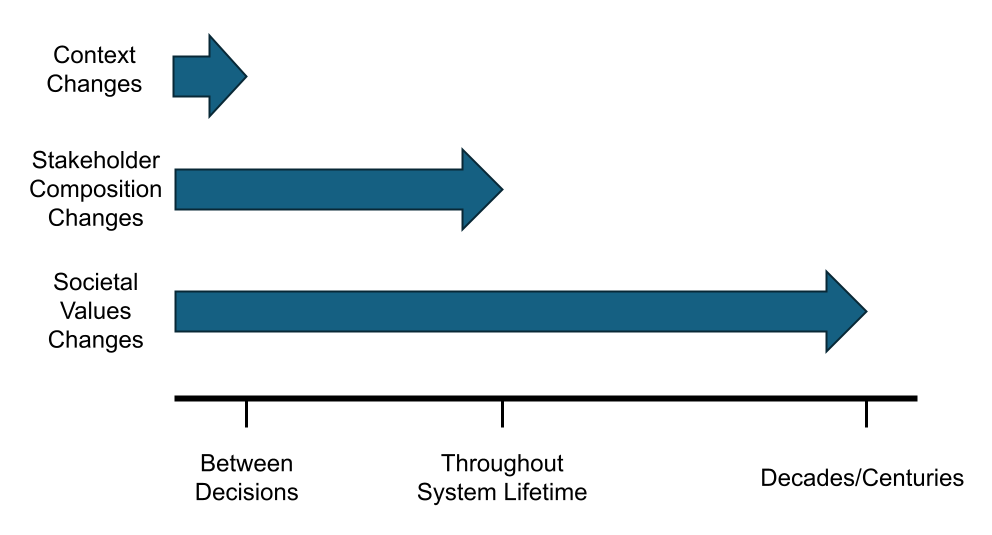}
    \caption{Operation time frames (x-axis) for the causes of changes in a system's relevant values and how they are prioritised (y-axis). Contextual changes occur the most frequently in decision-making scenarios faced by an autonomous agent, potentially as often as between individual decisions. Given that the way values are interpreted is determined by the context, a contextual change can have significant implications for value alignment. Stakeholder composition changes, in the form of relevant stakeholders changing or existing stakeholders' values changing, occur more slowly, but assuming the stakeholders will be static from the beginning of a system's deployment through its lifetime is risky from an alignment perspective. Finally, the wider values of the society that a system is embedded in, and the implications of these values for a system, changes the most gradually. The largest implication for a system is the need for it to be replaceable when it is no longer appropriate from an alignment perspective, which means considering how embedded the system may become in wider processes during its operation.}
    \label{fig:vals_change_timeline}
    \Description[Values are dynamic because of context changes, stakeholder composition changes, and societal value changes.]{Value dynamism due to changes in context happens most rapidly, as fast as between individual decisions. Value changes due to stakeholder composition changes occur throughout the system's lifetime. Societal value changes occur over decades or centuries.}
\end{figure}

Values are formed by human stakeholders and then contextualised by the agent's operating conditions to impact its decision-making. In Fig.~\ref{fig:vals_change_timeline}, we illustrate how these factors cause \textit{value dynamism}: the changes in values over time and the timescales these changes occur at. This creates a need for value alignment to be able to react to these changes, rather than only aligning with an initial state of values.

As discussed in Section~\ref{sec:contextualisation}, contextualisation is a core process in how values are interpreted. Because contexts have varying levels of granularity and are defined by complex circumstances, context changes can rapidly occur during an agent's operation and, hence, change how values are interpreted. As illustrated by the top arrow in Fig.~\ref{fig:vals_change_timeline}, these sorts of value changes occur over the shortest time frames: context change can occur in as short a time as between individual decisions. For instance, what values impact a social care robot's interactions with a patient may need to be very different to those informing interactions with a care professional or family member. Said robot could need to switch behaviours in the time it takes to move between the rooms these individuals are in. The frequent and unpredictable nature of context changes, combined with the very granular and diverse nature of contexts, poses one of the most challenging aspects of reliable value alignment.

Stakeholder composition reflects the goals of a system's stakeholders and the means of aggregating stakeholder values. Stakeholder goals and values change over time, which may result in misalignment with autonomous agents, even if they were aligned previously \citep{chaturvedi_ai_2023}. For example, a user of social media may originally prioritise their entertainment, which the platform's recommender engine is aligned with by providing content the user finds entertaining. Later, the same user wishes to be more mindful about their use of social media and instead prioritise their mental health by not clicking through content carelessly, even if it would be entertaining. The recommender system must adjust to the user's new goals and priorities to remain aligned.

To handle this form of dynamism, a regular assessment of the system's understanding of its stakeholders is necessary. This again highlights the need raised in Section~\ref{sec:val_al_stake} to engage with stakeholders as active participants in the system, but raises additional questions: how regularly this assessment needs to occur is not clear, and concerns about the user's privacy and cognitive demands on them are highly relevant. Some insights could potentially be gleamed from the literature on norm emergence \citep{morris2019norm, gelfand2024norm}, given the role of norms as representing grouped values.

In the long term, the contextualisation of values changes over time because society evolves. \citet{han_aligning_2022} cites the example of men duelling in centuries prior, as this was the appropriate response to their honour being threatened, but such practice would be illegal in many countries these days. While we still understand the concept of honour as a value, we practice it differently in modern times. It would also be inappropriate to assume that this will not continue in the future, as new generations develop new thoughts on morality and norms. Hence, we should acknowledge that systems may need to adapt to changing values in a greater sense than reconsidering the stakeholders of concern or being contextually adaptive. 

That said, the length of time over which \textit{societal value changes} occurs seems to be quite long, in the span of years, decades, or even centuries, as we show in the bottom and final arrow of Fig.~\ref{fig:vals_change_timeline}. The nature of systems, autonomous agents, and our concept of value alignment will no doubt further evolve in that time. Good practice for this kind of misalignment is to enable the evolution, or if necessary replacement, of systems using autonomous agents when needed, and avoiding them becoming too ingrained into any particular process such that these changes prove problematic. While this cultural shifting of values may not seem like a significant risk factor, the abundance of legacy systems employed in modern society cautions us not to take the ability to swap out a system for granted, particularly as that system's capabilities become more impactful. 

\subsubsection{Value Aggregation}
\label{sec:val_agg}

When inevitably dealing with multiple stakeholders, \textit{value aggregation}, the reconciliation of diverse stakeholder interests to determine system objectives and considerations, becomes necessary. We see this happening through the formation of ethic systems, as we discussed in Section~\ref{sec:impl_ethic_theo}, but aggregation also needs to occur on a smaller scale between individuals or smaller groups where cooperation is necessary for alignment.

While examining the literature, the majority of papers only looked at what \citet{liscio_value_2023} refers to as \textit{micro value alignment}, which is alignment occurring between two agents. The authors define \textit{macro value alignment} as leading to alignment between more than two agents, making use of mechanisms like norms. Particularly egregious was a lack of attempts to align multiple utility functions, given their prominence as a value encoding mechanism.

Since plenty of research exists on norm construction in multi-agent systems, a macro alignment mechanism according to \citet{liscio_value_2023}, then it is also possible that this work is not being properly integrated into the value alignment literature. This could be because it is not referred to as value alignment, or because it is not based on neural systems and hence not considered by researchers defaulting to a machine learning approach. Given the extensive amount of research in this norm-construction space, integrating this work into promising value alignment approaches would be an effective research direction, in line with what \citet{allen_artificial_2005} described as hybrid systems.

However, values vary enormously between stakeholders and often in incompatible ways \citep{butlin_ai_2021}. This produces what \citet{haas_moral_2020} refers to as a moving target for aggregation. Combining the diverse interests of stakeholders becomes more challenging as the range of relevant values increases, and fully satisfying everyone is often impossible in many situations. This is particularly true in the presence of conflicting values and their associated goals, which is often the case. For example, system providers may be able to create a more optimised system with more data and, hence, generate more profit, but users want their privacy to be respected. Social choice scholars such as \citet{arrow2012social} have written at length about the various impossibilities in achieving a complete satisfaction of all values between stakeholders, particularly if a utility-based approach is adopted \citep{aliman_augmented_2019}. 

This indicates that value aggregation is another value-laden process, much like selecting an ethical system to begin with. Trade-offs are a natural part of decision-making, particularly in the Schwartz model where some value categories, such as security and self-direction, are inherently opposed. However, trade-offs must still be decided regarding whose and which values are prioritised in a given context. When aggregating these different values for the sake of decision-making, it is plausible that different value aggregation methods will benefit different stakeholders and values in different ways. The more optimised a system becomes for one set a values, and hence the more trade-offs made, the more risk there is of not satisfying another set of values. This could pose serious risks if the trade-offs made are not carefully monitored, such as trading one group's safety or privacy for another group's profit or power.

In the literature, autonomous agents are often treated as having the single objective of aligning with a human. In practice, both autonomous agents and humans may start with pre-conceived goals and embedded values in a given scenario, even if it was not the designer's intention \citep{van_de_poel_embedding_2020}. Aggregation in this case is likely to require adjusting behaviour to accommodate another agent's values, so that the agent does not lose its original purpose, rather than replicate a different set of values completely. 

However, the risk of disempowering certain values mentioned in Section~\ref{sec:motivations} is relevant here. Proper alignment should not just involve one agent's value system being subsumed by another's. It is not as simple as saying that autonomous agents should align with humans rather than the other way around, as autonomous agents are almost certainly acting as a proxy for another human or group of humans. A careful approach to understanding the context behind potential value disempowerment in aggregation, with appropriate compromises as a result, is necessary for ethical aggregation, and further examination of value compromise in other domains can lend insight here.

Disempowerment of values in value alignment may occur through combative AI whose goals differ from human stakeholders, leading to competition in attempts to achieve different world states that satisfy different values \citep{nikolaidis_human-robot_2017, atkinson_value_2016}. This is an example of what happens when alignment fails \citep{chaturvedi_ai_2023}. As \citet{hadfield-menell_incomplete_2019} described it, agents pursuing different rewards become engaged in a strategic game against each other to try and achieve their own goals, even if it reduces the gains for the other agent, as in the classic prisoner's dilemma \citep{rapoport1965prisoner}.

The main form of combative AI we observed in our analysis, apart from discussion around manipulative superintelligences \citep{bostrom2017superintelligence, li_problems_2021, murray_stoic_2017}, was the case of autonomous agents overriding human decision-making. While the notion of surrendering our autonomy to autonomous agents may seem undesirable, research by \citet{milli_should_2017} suggested that there may be a level of nuance to this. In the case of humans expressing orders imperfectly, their study demonstrated superior performance of autonomous agents that attempted to support the human's underlying preferences. The research suggests that actions that may be seen as combative may have a time and place, but this still requires successful identification of a stakeholders' underlying values and careful ethical consideration of whether overriding is appropriate.

Transparency around the process of aggregation will also be essential. Given the need for value-based decisions about aggregation methods, different approaches may lead to different versions of alignment for the system, and different requirements to achieve alignment. Transparency will be necessary to mitigate the political risk this presents, from empowered stakeholders like designers and owners having their values overrepresented compared to less powerful stakeholders like users. Given the need to adequately model stakeholder values in value-aligned system, this could present a novel opportunity to model what happens to these values during different aggregation mechanisms more explicitly. Both developing methods for aggregation, and understanding how they impact different stakeholders, would be an intriguing line of research.

Concluding this section, we have drawn attention to the key factors when trying to use values to promote aligned decision-making between humans and autonomous agents. It is important to understand that values are not just static objectives that can be identified once and forgotten about, but dynamic mechanisms for approval or disapproval of a situation, obscured by their latent nature and the complex interactions between values and context. Furthermore, the need to aggregate values for satisfying groups of stakeholders creates methodological and political opportunities and challenges for research. There is still a great deal of work to be done in modelling values properly in autonomous systems.

\subsection{The Value Alignment Process}

So far, we have discussed the themes that illustrate what is driving value alignment as a relevant field of research, the particular challenges it faces, and the mechanics of its core concept, values. We now focus on how the literature portrays the process of value alignment in systems. Our themes of \textit{cognitive processes} and \textit{human-agent teaming} together illustrate how value alignment occurs at the systematic level.

\begin{figure}[ht!]
    \centering
    \includegraphics[width=\textwidth]{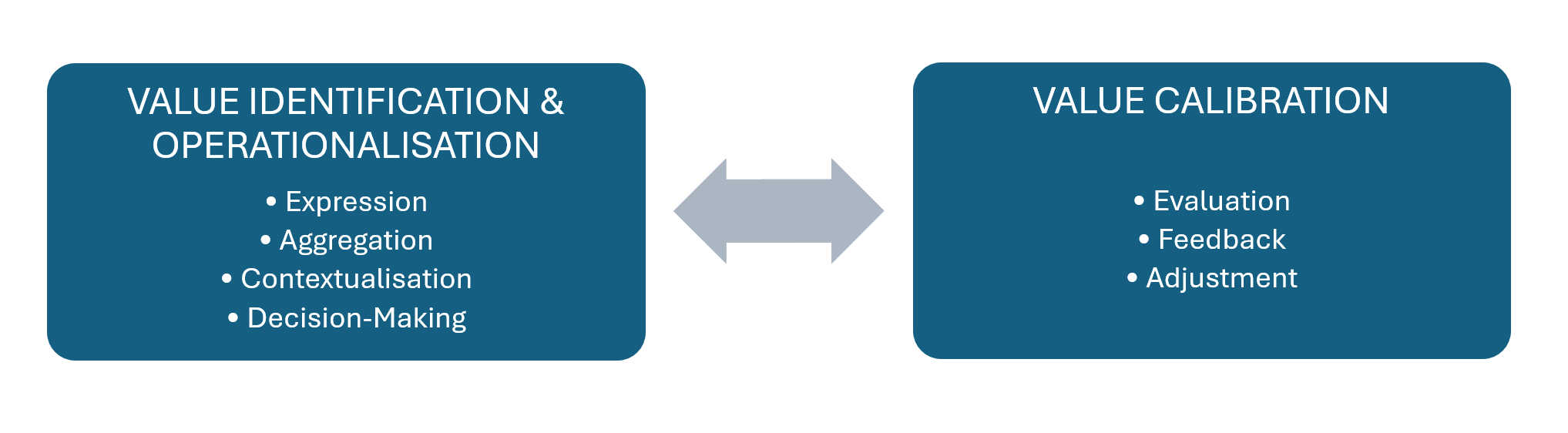}
    \caption{Diagram of the value alignment process and its two groups of subprocesses: value identification and operationalisation, and value calibration. Progression through the value alignment process between these groups is non-linear: outcomes in the calibration stage in the face of value dynamism may indicate a need to reconsider what values have been identified, or how they are operationalised.}
    \label{fig:val_al_process}
    \Description[The value alignment process is grouped into two categories: value identification and operationalisation; and value calibration.]{The value alignment process is grouped into two categories. Value identification and operationalisation includes the sub-processes of expression, aggregation, contextualisation, and decision-making. Value calibration includes evaluation, feedback, and adjustment. Systems can move back and forth between these groups of processes during their operation.}
\end{figure}

Analysis of the literature and our extracted codes led to the development of a concept diagram for how the subprocesses of value alignment interact in the operation of a system. We show this in Figure.~\ref{fig:val_al_process}.

In the diagram, we split the subprocesses into two stages: \textit{value identification and operationalisation} and \textit{value calibration}. Identification and operationalisation is the process of value communication and negotiation, then making those values functional in the relevant contexts for the sake of decision-making. We group these two verbs into one grouping of processes as the two are intertwined: operationalisation cannot occur without identification, but operationalisation can also reveal how values have not been sufficiently expressed or understood.

Value calibration accounts for the dynamic nature of values that we explored in Section~\ref{sec:val_dyn}. Its processes enable the communication of misalignment between agents, and promotes actions to correct it. Without calibration, value-aligned systems will inevitably become misaligned.

Although a system's life cycle will almost certainly start in identification and operationalisation, value alignment can move back and forth between the two groups of processes. Much like how operationalisation can state that values have not been understood sufficiently, calibration can reveal neglected values or stakeholders, or the need to redesign a system to attain changed values. Once the system has evolved to reflect this, calibration can continue. Hence, the process of value alignment between humans and AI agents is iterative.

\subsubsection{Value Identification and Operationalisation}
Value identification and operationalisation includes the processes of: value expression (including goal and preference expression); value aggregation; contextualisation; and value-based decision-making. We have already discussed expression in Section~\ref{sec:Chall_in_ValAl} and aggregation and contextualisation in Section~\ref{sec:vals_in_ValAl}, so here we focus on implementing values in relation to cognitive processes and human-agent systems. Viewing identification and operationalisation as part of a joint process combines the technical and normative aspects of alignment mentioned in Section~\ref{sec:tech_and_norm}, by considering the means of selecting and implementing stakeholder values alongside the technical requirements in doing so. It also illustrates the close relationship between identifying the values of interest in a given context, and then contextualising them to make them actionable by autonomous agents within the same context.

\paragraph{Value Learning, Teaching and Reasoning}

Expressing our priorities, particularly in a dynamic system, implies a need for other agents to be able to learn these priorities. Within our literature survey, the most common source of value-learning for autonomous agents was humans. This primarily seemed to occur due to the popularity of cooperative inverse reinforcement learning \citep{hadfield-menell_cooperative_2016} as a value alignment mechanism, and indeed the method has evolved into an approach designated reinforcement learning from human feedback. A particularly modern example of this approach has been in the development of large language models, which are employed in a vast range of contexts with minimal oversight, in order to train these models to develop behaviour in line with our preferences \citep{bai2022training}.

While humans make a natural candidate for teaching autonomous agents human values, it has its challenges. Humans are cognitively limited in some ways compared to autonomous agents, given our need for rest and limited ability to certain types of information relative to autonomous agents. As we mentioned in Section \ref{sec:expressing_prios}, expressing goals is difficult for humans, which complicates teaching them to autonomous agents. This is accounted for by some modern approaches, which aim to learn human preferences as these are easier for humans to express \citep{christiano_deep_2017, zintgraf_ordered_2018}. Regardless, our cognitive limitations remain an important consideration in any human-driven teaching process. 

Discussion drew a distinction between theoretical reasoning- reasoning what is the case- and practical reasoning - reasoning what to do \citep{zurek_reasoning_2022}. When it came to interpreting value-based knowledge learned from humans, practical reasoning was considered by the papers in our survey more than theoretical reasoning. 

While discussion focused more around practical reasoning, theoretical reasoning becomes very important if we consider the suggestion in \citet{gabriel_artificial_2020}, that ``AI would have to be aligned with some set of beliefs about value, not with value itself.''.  This would suggest that, for successful alignment to occur, autonomous agents are able to evaluate their beliefs about how values are realised, incorporate uncertainty into the reasoning process, and determine if these beliefs are consistent with humans in the system in which they exist. We can link this to human decision-making, as this was another process used to consider autonomous agent decision-making. A prominent observation was that humans make ethical decisions with incomplete and imperfect information \citep{robinson_action_2023, cervantes_autonomous_2016}, and it is natural to assume that autonomous agents will have to make ethical decisions under the same circumstances. \citet{russell2019human, bogosian_implementation_2017} and \citet{eckersley_impossibility_2019} all supported the necessity of implementing uncertainty in autonomous agents' moral decision-making to achieve alignment.

An interesting aspect of the reasoning process about values is the idea of \textit{emotional intelligence} in AI. This includes both recognising emotions in humans as a form of feedback, and using their own form of emotions in their reasoning process. While \citet{constantinescu_can_2022} observes that emotionless AI would be advantaged in making decisions free from undesirable emotions, it would also limit their ability to act in a virtuous capacity. \citet{cervantes_autonomous_2016} also raises the need for emotions in effective decision-making, citing neuroscientific evidence of the role of emotions in social and ethical contexts, both highly relevant in value alignment.

\paragraph{Decision-Making with Values}

Given our tendency to anthropomorphise autonomous agents \citep{duffy2003anthropomorphism, salles2020anthropomorphism}, it would seem natural that discussion of human decision-making would appear in works examining decision-making frameworks for autonomous agents. \citet{butlin_ai_2021} suggests that there is an acceptance that humans and animals use multiple decision-making systems, and this includes both a model-free and model-based reinforcement learning style approach. As \citet{cervantes_autonomous_2016} observes, human decision-making happens in a continuous fashion, with the majority of it being unconscious. In the context of values, \citet{zurek_value-based_2021} mentions that humans do not compare individual values and how they will be promoted, but evaluates various values simultaneously and use this to evaluate options. While emphasis in value alignment tends to be on constructing effective decision-making for autonomous agents, understanding how humans make decisions is essential in considering how both types of agents will interact with each other in pursuing alignment.

While no perfect formalism for how humans reason with values exists, it can generally be understood that values impact decision-making through contextualisation of values and the internal ordering over the values \citep{serramia_qualitative_2020, liscio_value_2023}, in order to evaluate and choose from the available options \citep{szabo_understanding_2020}. As \citet{han_aligning_2022} states, humans feel a ``calling'' to realise positive values, an ``ought to do''. But as \citet{waser_designing_2015} points out, while these values have enabled our survival to modern times, a lack of explanatory power regarding how values function makes their use difficult in ethical dilemmas without clear value resolutions. Simply having agents learn latent human values and mimic them in decision-making may not be ideal for complex ethical dilemmas, if such dilemmas lack consensus on a reasonable outcome. Indeed, this lack of explanatory power connecting values to decision-making increases the risk associated with relying on autonomous agents learning our values and trying to emulate them, as their behaviour becomes less predictable as a result, as discussed in Section~\ref{sec:motivations}. However, given that by definition, we want autonomous agents to do the things we value, trying to remove values from the process of integrating AI agents into systems would seem self-defeating.

Naturally, as autonomous agents are making more decisions, we can expect them to face more ethical dilemmas as well. As \citet{gabriel_artificial_2020} discusses, there is an interpersonal dimension to these conflicts: what is value-aligned for one person won't necessarily be for others. This brings us back to the challenges of value aggregation raised in Section \ref{sec:val_agg}, and the associated impossibility theorems. Given the perpetual conflict between humans' values and our construction of ethical frameworks that attempt to resolve these dilemmas, the choice of conflict resolution approaches will inevitably embody at least one ethical framework. \citet{cervantes_artificial_2020} cites \citet{broeders2011should} in how individuals use the moral rule that has given them the most success in the past when making judgments in similar situations, which \citet{cervantes_artificial_2020} incorporates into their own framework. From a design decision, it may be practical to focus on developing autonomous agents to recognise these ethical dilemmas and defer to humans rather than rely on them to resolve them for us, if we wish to maintain agency over our morality. Empirically, we may need to experiment with autonomous agents using different dilemma resolution methods in different contexts to determine which will be palatable in practice, as theorising alone seems unlikely to lead to a resolution.

For autonomous agent decision-making, utility-based approaches to learning values and ethics were abundant in our survey. However, little reasoning was done over the learned utility functions beyond taking the action that maximises the utility score. While this is the norm in utilitarian decision-making, critics point out that such an approach can be disastrous when the utility score fails to accurately capture the underlying objective \citep{zhuang_consequences_2020, gabriel_artificial_2020}. \citet{murray_stoic_2017} proposes that such systems should integrate a form of temperance into their reasoning, subjecting their maximisation goals to constraints on their impact. This may also have useful implications for aggregating values between stakeholders, by incorporating agreed trade-offs between groups.

Other authors investigated utility with multiple dimensions, emulating value pluralism \citep{mason2006value}, the assumption that multiple, incommensurable values exist. As stated in \citet{szabo_integrating_2022}, this assumption of multiple values existing is standard in value-based reasoning, either due to the distinctness of values, or the impracticalities of attempting to collapse them into a single value. Some works that included this multi-objective approach to utility functions include \citet{vamplew_human-aligned_2018, rodriguez-soto_guaranteeing_2021, haas_moral_2020, peschl_moral_2022}. 

Some other examples of value-based reasoning for decision-making in the survey were \citet{atkinson_value_2016}, who demonstrated how value based argumentation could be used to select actions without assuming other agents' preferences; \citet{holgado-sanchez_admissible_2023}, who examined how different contextualisations of values would impact the range of admissible actions; and \citet{zurek_reasoning_2022}, who used principle-based reasoning to select goals based on the values to be promoted.

\subsubsection{Value Calibration}
Value calibration is centred around evaluation, feedback, and adjustment. Proposed previously by \citet{firt_calibrating_2023} in the dimension of trust, we extend this stage to be necessary to accommodate the dynamic nature of values and context, as well as to allow corrections for the inevitable noise generated by the difficult nature of value expression, learning and aggregation.

\paragraph{Feedback}
Throughout the learning processes observed in the survey, the concept of feedback was pervasive. Given the difficulty in expressing values we discussed in Section \ref{sec:expressing_prios}, and the dynamism of values discussed in Section \ref{sec:val_dyn}, it seems sensible to believe that attempting to align to a set of values at one point in time may fail to maintain alignment when inspected at a later point. This could be due to misrepresentation of the values in the first place, or changes in the target values over time. To accommodate this, a capability for an autonomous agent to adjust its values and priorities at a given moment in response to feedback, potentially needing to learn new value representations in the process, will be required for robust value-aligned systems. This could be considered one of the key problems facing value alignment.

In particular, when human teachers are engaged with the agent in providing this feedback, the teaching process also needs to support the human teacher understanding how their feedback has been interpreted by the autonomous agent. \citet{sanneman_transparent_2023} explores this area through the use of explainable AI. Humans are also fallible, and may make teaching errors \citep{milli_should_2017} or teach undesirable values (relative to other stakeholders' values) to an autonomous agent \citep{gabriel_artificial_2020}. These issues reflect the need for transparency and explainability in the human-teaching process, both the values that have been learnt by the autonomous agent, for auditing and to support the human teacher, and the values held by human teacher, for the sake of auditing what values a system has been trained on.

Also relevant to the topic of feedback is the role of self-reflection in autonomous agents, as a form of internally-generated feedback. \citet{murray_stoic_2017} discusses this in the context of stoic ethics, where self-reflection would be used by autonomous agents to evaluate whether its past behaviour was aligned with its desired behaviour, while regret could be used in cases with no good options, to drive the agent to seek alternatives in similar future scenarios. \citet{shaw_towards_2018} considers self-reflection in a more normative sense, where it could be used to flag inconsistencies in an autonomous agent's learned principles. \citet{stenseke_artificial_2023} also poses self-reflection through a reflective and proactive method, with the former evaluating past behaviour similar to the work in \citet{murray_stoic_2017}, while the latter would allow the agent to simulate future scenarios to evaluate its potential behaviour. While self-reflection as a mechanic remains conceptual, it offers exciting potential for the value alignment process in the face of shifting and obscured alignment needs.

The idea of emotional intelligence in autonomous agents is again relevant here, as emotions can serve as a form of feedback that can signal a need for an agent to adjust its behaviour  \citep{salloum2025emotion, martinez2005emotions}. Emotions have been suggested as useful for inferring non-observable mental states \citep{tzeng_engineering_2022}, which can be used for triggering responses by the agent \citep{harland_ai_2023}, as well as for acting believably in human-computer interactions \citep{de_carolis_enhancing_2010} while avoiding acting in a manipulative way \citep{murray_stoic_2017, scheutz201113}. \citet{han_aligning_2022} also suggests that a need to feel emotions would be required for an autonomous agent to become independently value-aligned in a material value-ethics paradigm, where knowledge of values is only conveyed through emotions. Given their ability to contribute to reasoning about and learning values, the study of emotions in autonomous agents is at present an unexpected but intriguing avenue for value alignment work.

\paragraph{Evaluation}
We round out our analysis by evaluating whether systems have actually achieved value alignment or not. Given the challenges in modelling values that we have outlined up to this point, this is naturally a difficult assessment to make.

Our analysis agreed with the observation in \citet{feffer_preference_2023} that most evaluations in the research space rely on simulations or theoretical proofs. We were concerned by the lack of uniformity in the scenarios used for testing: authors would usually create novel simulations or repurpose their earlier work when testing their methodologies. Absent from the literature were reliable baseline scenarios which would support the comparison of different approaches to value alignment. 

While the trolley problem, a well-explored ethical dilemma example in moral philosophy \citep{thomson1984trolley}, was discussed in some studies \citep{faulhaber_human_2019, peterson_value_2019, loreggia_metric_2019}, \citet{eckersley_impossibility_2019} is quick to point out the negligible relevance of the trolley problem to the space of possible scenarios an agent might have to consider in value alignment. We have illustrated other significant challenges in value alignment, such as goal expression and context modelling, to which the trolley problem adds little insight. Our conclusion is that the trolley problem could prove a useful baseline for developing theoretical reasoning about values in autonomous agents, given that its well-studied nature provides complimentary thought on any produced reasoning. However, authors should be wary of confirmation bias when using the trolley problem and ideally test their model on alternative value-based scenarios.

We theorise that there are two causes for the lack of baseline scenarios in the surveyed literature. First, ambiguity on what value alignment actually is makes it difficult for authors to assess whether scenarios from other research would be suited for their own interpretation of value alignment. The goal of this paper is to help alleviate this by working towards unifying the nature of the topic, while also identifying sub-problems for which tractable baselines could be generated. The ability to reproduce a human's identified values, obtained through methods from psychology, through agent-interaction, would be one example baseline. Successful context modelling, and hence policy switching, could be another - one we already see studied through explainable AI approaches to domains like self-driving cars \citep{atakishiyev2024explainable}.

Second, it is difficult to envision scenarios where different values and goals can be clearly articulated, simple enough to implement and test repeatedly while also avoiding overly simple behaviour that has an ``obviously correct'' answer. For example, \citet{rodriguez-soto_instilling_2022} assesses their value alignment methodology through their public civility game, where an agent must learn to throw a rubbish bag to the bin rather than in its fellow agent's path. While this is fine for confirming that their agent has learned desirable behaviour, there is no real alternative behaviour here that could also be considered aligned from a different agent's perspective. Someone may be in favour of throwing trash in their colleague's path, but attempts to justify this would likely introduce new values such as this being the more efficient option, or simply preferring to be mean-spirited. There is a difficult trade-off in designing scenarios between the necessary complexity of behaviour and ease of implementation of test scenarios: additional complexity increases the difficulty in constructing such a scenario, as well as measuring the effect of different factors on outcomes. The development of these baseline test scenarios, and the demonstration of their fitness for purpose, would be a valuable contribution to value alignment research.

Assessing value alignment is not a one-off exercise but rather an ongoing process. As we have established in Section~\ref{sec:val_dyn}, values change over time through contexts and stakeholders changing. And as \citet{van_de_poel_embedding_2020} points out, the evolution of systems of autonomous agents can lead to the embedding or dis-embedding of different values over time as the agents adapt within the system. Questions regarding the frequency of this assessment in the face of dynamic values, and how to respond when misalignment is detected, remain open.

Finally, we emphasise the point made previously by \citet{sanneman_validating_2023} that there is a significant shortage of empirical testing of value alignment systems, including both humans and autonomous agents. Simulations and theoretical approaches are useful for testing methodologies or ideas initially. However, our interest in value alignment comes from the need to have humans and autonomous agents interact, and we have established that values in practice are highly dynamic and interactions with them are difficult to predict. This is made more difficult by the cognitive hurdles in interpreting values and validating that alignment has happened. Research requires contextual interaction with human stakeholders to generate data that can be placed in the context of the system's intended operating environment, and the values and contextualisations these imply. Unfortunately, this is difficult to implement for many systems, given that risking being misaligned with stakeholder values in many scenarios can cause harm, be it physical, mental, or otherwise.  

While studies making use of humans were encountered in our survey \citep{liscio_what_2022, siebert_estimating_2022, svegliato_ethically_2020, zintgraf_ordered_2018, ficici_simultaneously_2008}, the focus of these was on modelling user values and preferences. This is an essential step to value alignment, but it only approaches the value identification stage of the process. Assessment of a system's alignment with a stakeholder's values will require some ability to assess system alignment in its operating contexts, and a means to validate this alignment for auditors. This brings us back to the need for formalised mechanisms for evaluating alignment.

Two studies did investigate humans exploring components of the system, or interacting with the system to assess alignment. \citet{sanneman_validating_2023} used latent factor analysis to investigate alignment between stakeholder goals and agent reward function, identifying two latent values they labelled 'feature alignment' and 'policy alignment'. In our model, we would describe this as the need to represent values appropriately for stakeholders, and prioritise different values via proxies in an agreeable manner. \citet{nikolaidis_human-robot_2017} investigated the performance of human-robot teams cooperating in situations where the robot could override human preferences. It was observed that an adaptable approach to the human's degree of cooperation improved performance relative to pure obedience. While the lens of this study was on user trust rather than value alignment, it still provides insight into the nature of humans interacting with autonomous agents and how their goals, which are derived from values, can be supported through the nature of said interaction. 

While theory development and simulations are essential tools in the development of value-aligned systems, studies like the previous two demonstrate the benefits and needs for more empirical research. Given the subjective nature of values and relatively limited experience of humans interacting with autonomous agents, involving humans in more stages of both researching and developing value-aligned systems is vital to advance our understanding about the possibilities and limitations of value alignment and how these can be achieved.

In conclusion, we see that the value alignment process is a dynamic human-AI process, and a hugely complex one at that. Achieving alignment between humans and autonomous agents in systems will require more than just teaching autonomous agents to learn a set of values: it will require understanding how these values are communicated and modelled by both agent types and how feedback can be used to adjust them as necessary. In addition to the essential need to modelling context appropriately, other forms of information like emotion can provide insight in the alignment process, Finally, understanding how the interactions between both humans and autonomous agents impacts alignment and assessing the state of alignment in the face of these ongoing interactions is not currently well understood, but this needs to change for robust alignment to be achieved.

\section{Discussion}
\label{sec:discussion}

Our objective for this paper was to gain a deeper understanding of the value alignment challenge for humans and autonomous agents. From our analysis we can define value alignment as an ongoing dynamic process of identifying, operationalising and calibrating values, that is complicated by the abstract nature of values and contextualization, the difficulties in identifying and communicating values between humans and autonomous agents and evaluating the state of values, accommodating the dynamic nature of values, and the ethical and political risks design decisions around values and their aggregation entails.

From our analysis, it is clear that value alignment is a complex topic. It cannot be described in terms of simple objectives, as the concepts of `value' and `alignment' are themselves inherently complex and subject to interpretation. The scope of processes involved in value alignment and the range of subjects that can contribute towards understanding it is broad. Effectively formalising value alignment as a process that can be implemented reliably will require a concerted interdisciplinary effort to reach viable solutions and should not be viewed only as a computer science concern. 

Furthermore, researchers will benefit from narrowing their focus to particular aspects of the value alignment process, such as value representation or managing uncertainty in environmental perception. In doing so, they should not lose sight of the wider elements of value alignment and how their efforts may impact upstream or downstream processes, given the ongoing nature of value alignment as a process.

Value alignment should also be understood as a process of human-agent interaction, as we see through the need to calibrate alignment through teaching values and feedback between humans and autonomous agents. Attempting to approach value alignment while focusing only on AI mechanisms, under the assumption that humans can be integrated later, risks starting with a false premise. Ignoring the humans, the sources of values and the drivers of their dynamic and complex nature, in the process of value-alignment is difficult if not outright naive.

We emphasise that value alignment is an iterative process. We have made clear the dynamic nature of values, being highly sensitive to context and stakeholders, and these will change throughout the system's operation. This makes alignment inherently unstable, and even if the first version of an agent deployed and evaluated appears appropriately aligned, this is no guarantee of future alignment. This iterative nature emphasises the role of online learning and self-maintenance of autonomous agents in value alignment contexts.

Regular evaluation and adaptability are essential to maintain robust value-aligned systems. Interaction between humans and autonomous agents to produce feedback is also key to reaching an aligned state, through communicating states of misalignment and providing guidance to correct it. Not only will this enable alignment, it also has scope for improving user trust towards such systems and supporting the appropriate adoption of autonomous agents in society.

It is clear from our analysis that value-aligned systems need more empirical research if the field wants to continue advancing. There are numerous challenges in successful expression of human values across diverse contexts, as well as too many uncertainties in how humans will respond to attempts to embed these values in autonomous agents or the impact these agents will have on human values. These cannot be anticipated purely from theoretical or simulation approaches alone. 

This should not be taken as an indication that theory and simulation have no use, as designing autonomous systems is a lengthy and costly process, and attempting to start with humans involved will inevitably add hurdles in undertaking the research. We instead believe that we are in a position now where empirical data on the value alignment process can add value, and researchers should embrace it. Collecting this empirical data can naturally include risks when exposing humans to autonomous agents, so care must be a priority. Researchers with experience in human trials can easily lend expertise here.

The fact that value alignment is difficult should not be understated. It is very doubtful that a single mechanism can overcome the numerous challenges presented by value alignment. Instead, a combination of components external and internal to the autonomous agent undergoing alignment will likely be needed. Overall alignment will also require regular monitoring during the system's lifetime, to promote calibration, and mechanisms to enable coordination between both human and autonomous agents. With this in mind, it is important to build resilience into a system for when misalignment inevitably does occur among these moving parts.

Finally, it is important to note that alignment should not be expected to be achieved with all potential stakeholders, but this can be acceptable. A system that prejudices a certain group towards more severe jail sentences would be misaligned with that group, and this would not be acceptable as the harm to their values of justice and freedom would be severe. In comparison, a generative art model that produces art that is not to some users tastes would not be aligned with them, but this would be fine as the system does not negatively impact their values otherwise, and alternatives may exist that can replace the given model. Deciding when misalignment is acceptable is itself a value-laden decision, and another part of the complex process of designing value-aligned systems.

\subsection{Opportunities and Future Directions}

Given its nature as a complex, interdisciplinary topic, there is much value to be gained from other disciplines' insights into the core challenges we have identified in this paper. Fig~\ref{fig:ValAl_subjects} illustrates the subjects encountered in this analysis, but it is not exhaustive with regard to what disciplines can contribute. For example, human-computer interaction as a research topic could be critical in the calibration group of processes, as the field could offer insight in how to engage stakeholders interacting with autonomous agents to assess whether they feel the system is aligned or not, or develop better methods for them to use in reporting their values to autonomous agents. Reducing barriers for interdisciplinary knowledge sharing and collaboration in value alignment research will be very useful.

Tackling one of the main challenges in value alignment, a better understanding of how values and goals are expressed, and where mistakes in this process occur, would support alignment. Given the cognitive difficulties involved in expressing these concepts, attempts to formalise some processes in the context of aligning autonomous agents could potentially control for these challenges and reduce errors in identification.

Alternatives to utility function, which, as discussed in Section~\ref{sec:expressing_prios}, are prone to under-specifying objectives or integrating more refined forms of utility functions into value alignment, are also due investigation. Other approaches to modelling values entirely, such as attempts to encode virtues or deontic systems, may be necessary for achieving alignment, but these are still works in progress.

Value aggregation is also in need of further investigation. The current norm is aligning two agents, usually at least one of which has a static policy. Pushing for research that attempts to align more than two agents using utility functions would be a relatively low-effort but simple starting challenge, especially given the existing work on using norms for multi-agent systems, and some of this research no doubt already exists in the multi-agent or swarm AI space. Potentially of greater interest would be examining different methods of alignment and their impacts on the form alignment takes, particularly in what stakeholders are disadvantaged or empowered. Such results would not only be useful in support design decisions, but also in understanding the political implications of autonomous agents as a technology and how these can be accounted for.

The contextualisation of values as a process is currently vague, but it is critical to value alignment. Understanding and formalising this process, even if it is only in the context of aligning autonomous agents, could add immense value to the topic and improve transparency in design. Case studies of how values have already been embedded in autonomous agents through processes like value-sensitive design \citep{friedman_value-sensitive_1996} and how these relate to value alignment would be particularly useful in generating empirical results.

Core to modelling contextualisation is the need for means to model context and how it is recognised by autonomous agents. This would support the explainability of agent decision-making by indicating the state features that describe a given context, possibly giving a compressed representation of all state features. If this can further be enhanced by understanding how contexts affect values specifically, then this would further improve the transparency of decision-making in autonomous agents. This is no straightforward task, but it would add necessary value in supporting value alignment, and is worth our continued attention.

Value calibration was neglected in the research we analysed, but given the dynamic nature of values, this needs to change. Ways to effectively track stakeholder values in dynamic situations are a component of this. If it quickly becomes apparent on deployment that misalignment is an issue, this needs to be identified before negative outcomes occur. This links into the need for understanding how often evaluation and calibration need to occur; which itself points to a need for a better understanding of value dynamism. In addition, further development of the methodologies for engaging stakeholders and assessing their alignment with active systems will be critical to effective calibration.

Testing approaches to value alignment is one last area that would strongly benefit from attention and unification. The current environment of individual experimental designs, often lacking in complexity or validation, limits the ability to say whether an approach is fit for purpose as it prevents comparison with other methods and obscures the complexity of values and context. Resolving this will require focused attention on what makes a good value alignment test scenario, and to what extent these can be generalised between environments to produce benchmarks. This is also another area which invites empirical research, and expertise from those with experience in measuring human-computer interaction.

\subsection{Limitations}

As stated in Section.~\ref{sec:methodology}, we limited the content of our survey to the English-language papers due to a lack of translation capability. We also observed that non-Western ethical systems and values were neglected in the value alignment research that we analysed. Given that this paper was authored by a Western team, this would have further compounded the effect of Western values on the interpretation of the data. 

As a result, this paper presents a dominantly Western perspective on the process of value alignment. This is not ideal, as value alignment will be required in technologies around the globe and to process values in numerous regions, not just the West. Future studies should keep this in mind and try to bridge this gap. Complimenting our understanding of value alignment with further research on non-Western approaches could offer insights into geographic differences in the value alignment process and challenges that may arise in the face of globalised AI technology.

Another limitation of our analysis is our use of peer-reviewed academic literature from Scopus to source our data. Scopus does not include content from most workshops and a significant number of conferences in computer science. While our bibliography search extended our paper list, it still relied on relevant papers being cited in Scopus-sourced papers to begin with. The fact that we excluded papers published after 2023 adds to this concern, given the increasing rate of studies emerging, as demonstrated in Fig.~\ref{fig:StudiesByTheme_timeline}. Future studies can replicate our methodology, however, to extend our analysis with these more recent discussions.

While we analysed academic literature, value alignment is also a popular topic of discussion on non-academic platforms frequented by industry practitioners and non-academic thinkers, and there is a significant body of writing on these platforms. This content is potentially not subject to the same scrutiny as academic literature, which is why we did not include it in our review, but it may still offer insight in future reviews. Even if it does not provide any new insights into the value alignment process, it would give an insight into the thought processes of the people doing value alignment in many modern AI systems.

\section{Conclusion}
\label{sec:conclusion}

The goal of this paper was to review different perspectives in the literature in order to understand value alignment through thematic analysis. Our research has led us to define value alignment as a complex, interdisciplinary topic with an interconnected set of themes. These themes lead us to the idea that value alignment is about enabling humans and autonomous agents to interact in ways that support multiple competing dynamic human values, which are highly sensitive to the operating context and constrained by expression challenges. The numerous challenges emerge from how these values are identified and operationalised in ways compatible with both humans and autonomous agents across multiple contexts, and then how alignment in terms of these values is calibrated throughout the system's lifetime in the face of political risks and ethical disagreement.

In summary, these are our key observations from this survey:

\begin{itemize}
    \item Value alignment is \textit{complex}. Value alignment as a process cannot be described in terms of simple objectives. The scope of processes involved in value alignment, and the range of subjects that can contribute towards understanding it, is broad. The process will require a concerted interdisciplinary effort to reach viable solutions, and should not be viewed only as a computer science concern. Researchers will also benefit on recognising the subproblems in value alignment requiring attention, such as value modelling or alignment evaluation, rather than viewing value alignment as a single monolithic problem.
    \item Value alignment is a process of \textit{human-machine interaction}. Attempting to approach it while focusing only on the technical or normative dimensions risks starting with a false premise, due to the interconnected nature of the social and technical dimensions in developing value-aligned systems. Successful research and development of value-aligned systems needs to move beyond theory and simulations to include more empirical research including humans, or at least an understanding of human-agent interaction, if the field wants to continue advancing.
    \item Value alignment is \textit{iterative}. Values are highly sensitive to context, and operating contexts will change repeatedly throughout the system's lifetime. Even if the first version of an agent deployed is appropriately aligned, this will not last forever. Adaptability is essential to maintain robust value-aligned systems. Interaction between humans and autonomous agents is also key to reaching an aligned state, through communicating states of misalignment and providing guidance to correct it. Not only will this enhance alignment, it will also improve user trust towards the system and support the adoption of autonomous agents by society.
    \item Value alignment is \textit{two-way}. Humans and autonomous agents will adapt to each other. While autonomous agents will ideally learn to embody human values, there are risks around humans deviating away from their own ideal values through interaction with autonomous agents. To avoid this leading to undesirable value changes, value checks should be available for both humans and autonomous agents engaging with a system.  
    \item Value alignment is \textit{difficult}: the abstract and dynamic nature of values combined with the sensitivity of the contextualisation process makes value alignment a very unstable process. It is very doubtful that a single mechanism can resolve the diverse issues in value alignment; instead, a combination of components external and internal to the autonomous agent undergoing alignment will be needed, and overall alignment will require regular monitoring during the system's lifetime. With this in mind, it is important to build resilience into a system for when misalignment inevitably does occur.
\end{itemize}

As autonomous agents become more embedded in our society, effective value alignment will be crucial to ensure these agents behave in ways that support our goals rather than oppose them, and this includes robust methods for assessing when misalignment is occurring for each of the system's stakeholders. By drawing attention to the many aspects of this process, we hope that this paper will support future researchers wanting to engage in the field by clarifying the research opportunities available to them, as well as flagging potential risks that should be considered in developing value-aligned systems. 

\begin{acks}
    This work is supported by the UK Research and Innovation under Grant No.: EP/S023437/1.
\end{acks}

\section*{Open Source Statement}
The data used in this paper will be made publicly available at \url{https://github.com/JamMack/Understanding-Value-Alignment-as-a-Process-a-Survey?tab=readme-ov-file} upon its publication.


\appendix
\section{Search Details}
\label{sec:search_details}
Our database of papers from \textit{Scopus} was initially constructed from three searches. We selected papers to include in the review by first by examining the title and abstract, and then by examining the full text. Further papers were obtained by examining the bibliographies of papers included in the first coding pass.

\subsection{Initial Value Alignment Search}

\textbf{Search Terms:}\\
TITLE-ABS-KEY ( "Value Align*" OR "Alignment Problem*" OR "Value-based reasoning" OR "Human Preference*" ) AND TITLE-ABS-KEY ( "Artificial Intelligence*" OR "AI*" OR "Agent" OR "Autonomous" OR "Intelligent System" ) AND NOT ( "time-series" OR "knowledge-graph" OR "SINS" OR "protein sequence*" OR "genome*" ) AND PUBYEAR $>$ 1900 AND PUBYEAR $<$ 2024 AND ( LIMIT-TO ( SUBJAREA , "COMP" ) ) AND ( LIMIT-TO ( PUBSTAGE , "final" ) ) AND ( LIMIT-TO ( LANGUAGE , "English" ) )

\textbf{Date of Search:} November 7th 2023 \\
\textbf{Initial Results:} 535 \\
\textbf{Included results after abstract \& title check:} 162 \\
\textbf{Included results after full text check:} 90 \\

Our first set of TITLE-ABS-KEY terms aimed to identify value alignment related papers, or those that connect to human preferences. Our second set of TITLE-ABS-KEY terms restricts this to papers discussing artificial intelligence related systems. Our excluding TITLE-ABS-KEY terms are based on a previous scoping review we conducted. We used them to exclude irrelevant topics that trigger based on our searching for alignment in computer science. We then filter by publication year and subject area, and the last two criteria restrict our search to papers that have been quality assessed through being at final publication stage, and will be in a language we can work with.

\subsection{Virtue Ethics and Social Contracts Search}

\textbf{Search Terms:}\\
TITLE-ABS-KEY ( "Virtue ethic*" OR "Social contract" ) AND TITLE-ABS-KEY ( "Artificial Intelligence*" OR "AI*" OR "Agent" OR "Autonomous" OR "Intelligent System" ) AND NOT ( "time-series" OR "knowledge-graph" OR "SINS" OR "protein sequence*" OR "genome*" ) AND PUBYEAR $>$ 1900 AND PUBYEAR $<$ 2024 AND ( LIMIT-TO ( SUBJAREA , "COMP" ) ) AND ( LIMIT-TO ( PUBSTAGE , "final" ) ) AND ( LIMIT-TO ( LANGUAGE , "English" ) )

\textbf{Date of Search:} November 9th 2023 \\
\textbf{Initial Results:} 138 \\
\textbf{Included results after abstract \& title check:} 46 \\
\textbf{Included results after full text check:} 31 \\

Our second search replaced the first set of TITLE-ABS-KEY terms in order to ensure coverage of virtue ethics and the social contract, as these were deemed relevant to the value alignment problem. The rest of the search followed the same reasoning as the first search.

\subsection{Multi-Agent Systems Search}

\textbf{Search Terms:}\\
TITLE-ABS-KEY ( "Value Align*" OR "Alignment Problem*" OR "Value-based reasoning" OR "Human Preference*" ) AND TITLE-ABS-KEY ( "Multi Agent System" OR "Agent Based" ) AND NOT ( "time-series" OR "knowledge-graph" OR "SINS" OR "protein sequence*" OR "genome*" ) AND PUBYEAR $>$ 1900 AND PUBYEAR $<$ 2024 AND ( LIMIT-TO ( SUBJAREA , "COMP" ) ) AND ( LIMIT-TO ( PUBSTAGE , "final" ) ) AND ( LIMIT-TO ( LANGUAGE , "English" ) )

\textbf{Date of Search:} February 2nd 2024 \\
\textbf{Initial Results:} 53

This third search was performed after the initial phases of coding indicated a strong motivation for considering the value alignment problem through a multi-agent system perspective. The purpose of this search was to ensure that we properly integrated thought from multi-agent systems into the review, and to avoid replicating existing work. This was again achieved by modifying the second set of TITLE-ABS-KEY terms, while keeping the rest of the search the same.

We only included 8 papers from this search, as the rest of the papers had either already been added from the previous two searches or were irrelevant.

\section{Inclusion/Exclusion Criteria}
\label{sec:incl_excl_crit}

\subsection{Abstract \& Title Check Criteria}
\textbf{Ex1\_Moral Values and Preferences}\\
\textit{Is the paper discussing values in the context of moral and social values, rather than only technical values (e.g. measurements and function outputs)?}

We defined moral/social values as factors that lead to preferences in humans, or otherwise help coordinate group decisions.

We did not require an attempt to define or measure these values to have been made in the paper.

The purpose of this criteria was to exclude papers that only used the term value in the sense of metrics, without looking at the type of values we were interested in for this survey.

\textbf{Ex2\_Values and Preferences Learning}\\
\textit{Is the paper discussing the challenge of integrating values, feasibly in the form of preferences, in forms of technology?}

The purpose of this criteria was to include papers that linked values to technology. Our inclusion of preferences as a viable expression of values was based on our prior knowledge before starting the survey.

\subsection{Full Text Check Criteria}
\textbf{In1\_Accessibility }\\
\textit{Do we have access to a copy of the paper?}

This included open-source access or institutional access. It did not include access that could be obtained through a separate purchase or similar mechanic.

\textbf{In2\_Value\_Alignment}\\
\textit{Does the paper discuss the challenge of getting humans and autonomous systems to act with aligned behavior?}

We defined aligned behavior here as behavior that is considered acceptable by humans according to their moral/social values. 

At this stage of the research we did not discriminate between alignment being targeted at a single human or multiple humans.

We allowed discussions of multiple agents without specifying which agents were human and which were autonomous, as long as there was no reason this could not include mixtures of autonomous agents and humans.

We excluded alignment discussed only in the context of physically coordinated actions, as was common in human-robot interaction papers.

\textbf{In3\_Definitions}\\
\textit{Does the paper include a definition of value alignment to some extent, even if it does not explicitly call it such?}

This referred to an attempt to explain the meaning of the value alignment process, as per our working understanding of it (humans and AI agents acting in ways that agree with each other), beyond simply indicating that it is a problem that exists.

We included indirect definition through proposing a methodology for solving the problem. We interpreted this as the proposed solution through this methodology indicating the criteria for solving value alignment, in whole or partially.

\subsection{Full Coding Pass Criteria}
\label{sec:incl_excl_crit_full_pass}

\textbf{Ex4\_Value Aligned AI Governance}\\
We did not include topics focusing on the governance around value aligned AI. This was outside the scope of this survey.

\textbf{Ex5\_Moral Status of AI}\\
We did not include papers focusing on the moral status or moral capabilities of AI. This was outside the scope of this survey.

\textbf{Incl6\_Implementation method OR Design Discussion OR VA Modelling}\\
\textit{Does the paper contain a theoretical/practical implementation of value alignment?}

\textit{OR}

\textit{Does the paper contain discussion around the design of value aligned systems, including overall structure or individual components?}

\textit{OR} 

\textit{Does this paper contain a discussed or implemented model of value aligned systems?}

The intent of this criteria was to select papers that included practical components of value alignment, or contributed discussion to how these  components should be designed.

\textbf{Ex7\_Specific Values}\\
We did not include papers that only focused on specific values, such as fairness or explainability. Our goal was to understand the process generally, rather than in the context of a specific value.

\textbf{In8\_Normative Behaviour}\\
We included papers that discussed social, normative or some other organisational alignment of agents. This was done to include papers focusing on normative modelling, as we considered them relevant to understanding the value alignment process.

\textbf{In\_Final}\\
\textit{Will we include the article in the review?}

We fully coded the paper if it passed all exclusion checks in the full coding pass criteria checks, and at least one of the inclusion checks in the full coding pass criteria.

\section{Reproducibility Checklist for JAIR}

Select the answers that apply to your research -- one per item. 

\subsection*{All articles:}

\begin{enumerate}
    \item All claims investigated in this work are clearly stated. 
    [yes]
    \item Clear explanations are given how the work reported substantiates the claims. 
    [yes]
    \item Limitations or technical assumptions are stated clearly and explicitly. 
    [yes]
    \item Conceptual outlines and/or pseudo-code descriptions of the AI methods introduced in this work are provided, and important implementation details are discussed. 
    [NA]
    \item 
    Motivation is provided for all design choices, including algorithms, implementation choices, parameters, data sets and experimental protocols beyond metrics.
    [yes]
\end{enumerate}

\subsection*{Articles containing theoretical contributions:}
Does this paper make theoretical contributions? 
[yes] 

If yes, please complete the list below.

\begin{enumerate}
    \item All assumptions and restrictions are stated clearly and formally. 
    [yes]
    \item All novel claims are stated formally (e.g., in theorem statements). 
    [yes]
    \item Proofs of all non-trivial claims are provided in sufficient detail to permit verification by readers with a reasonable degree of expertise (e.g., that expected from a PhD candidate in the same area of AI). [yes]
    \item
    Complex formalism, such as definitions or proofs, is motivated and explained clearly.
    [yes]
    \item 
    The use of mathematical notation and formalism serves the purpose of enhancing clarity and precision; gratuitous use of mathematical formalism (i.e., use that does not enhance clarity or precision) is avoided.
    [yes]
    \item 
    Appropriate citations are given for all non-trivial theoretical tools and techniques. 
    [yes]
\end{enumerate}

\subsection*{Articles reporting on computational experiments:}
Does this paper include computational experiments? [no]

\subsection*{Articles using data sets:}
Does this work rely on one or more data sets (possibly obtained from a benchmark generator or similar software artifact)? 
[yes]

If yes, please complete the list below.
\begin{enumerate}
    \item 
    All newly introduced data sets 
    are included in an online appendix 
    or will be made publicly available upon publication of the paper.
    The online appendix follows best practices for long-term accessibility with a license
    that allows free usage for research purposes.
    [yes]
    \item The newly introduced data set comes with a license that
    allows free usage for reproducibility purposes.
    [yes]
    \item The newly introduced data set comes with a license that
    allows free usage for research purposes in general.
    [yes]
    \item All data sets drawn from the literature or other public sources (potentially including authors' own previously published work) are accompanied by appropriate citations.
    [yes]
    \item All data sets drawn from the existing literature (potentially including authors’ own previously published work) are publicly available. [yes]
    \item All new data sets and data sets that are not publicly available are described in detail, including relevant statistics, the data collection process and annotation process if relevant.
    [NA]
    \item 
    All methods used for preprocessing, augmenting, batching or splitting data sets (e.g., in the context of hold-out or cross-validation)
    are described in detail. [NA]
\end{enumerate}

\end{document}